\documentclass[fleqn,useAMS,usenatbib]{mnras}
\usepackage{times}
\usepackage{rotating}
\usepackage{amsmath}
\usepackage{amssymb}
\usepackage[dvipsnames]{xcolor}
\usepackage{graphicx}
\usepackage{graphics}
\usepackage{hyperref}
\hypersetup{
	colorlinks   = true,
	citecolor    = blue
}

\def\apj{{ ApJ}}
\def\apjl{{ApJL}}

\def\aap{{ A\&A}}

\def\mnras{{ MNRAS}}

\def\nat {{ Nature}}

\def\prd{{ Phys. Rev. D}}

\def\pasa{{PASA}}

\long\def\symbolfootnote[#1]#2{\begingroup%
	\def\thefootnote{\fnsymbol{footnote}}\footnote[#1]{#2}\endgroup}

\newcommand{\gae}{\lower 2pt \hbox{$\, \buildrel {\scriptstyle >}\over {\scriptstyle \sim}\,$}}
\newcommand{\lae}{\lower 2pt \hbox{$\, \buildrel {\scriptstyle <}\over {\scriptstyle \sim}\,$}}
\newcommand{\aprop}{\lower 2pt \hbox{$\, \buildrel {\scriptstyle \propto}\over {\scriptstyle \sim}\,$}}

\newcommand\restr[2]{{
		\left.\kern-\nulldelimiterspace 
		#1 
		\vphantom{\big|} 
		\right|_{#2} 
}}

\def\n13{\mbox{\large n}_{_{13}}}

\begin{document}
	
	\title[FRB variability]
	{What does FRB light-curve variability tell us about the emission mechanism?}
	
	\author[Beniamini \& Kumar]{Paz Beniamini$^1$ $\&$  Pawan Kumar$^2$
		\\ $^1$Division of Physics, Mathematics and Astronomy, California Institute of Technology, Pasadena, CA 91125, USA \\
		$^{2}$Department of Astronomy, University of Texas at Austin, Austin, TX 78712, USA}
	
	\maketitle
	
	\begin{abstract}
		A few fast radio bursts' (FRBs) light-curves have exhibited large intrinsic modulations of their flux on extremely short ($t_{\rm r}\sim 10\mu$s) time scales, compared to pulse durations ($t_{\rm FRB}\sim1$ms). Light-curve variability timescales, the small ratio of rise time of the flux to pulse duration, and the spectro-temporal correlations in the data constrain the compactness of the source and the mechanism responsible for the powerful radio emission. The constraints are strongest when radiation is produced far ($\gtrsim 10^{10}$cm) from the compact object. We describe different physical set-ups that can account for the observed $t_{\rm r}/t_{\rm FRB}\ll 1$ despite having large emission radii. The result is either a significant reduction in the radio production efficiency or distinct light-curves features that could be searched for in observed data. For the same class of models, we also show that due to high-latitude emission, if a flux $f_1(\nu_1)$ is observed at $t_1$ then at a lower frequency $\nu_2<\nu_1$ the flux should be at least $(\nu_2/\nu_1)^2f_1$ at a slightly later time ($t_2=t_1\nu_1/\nu_2$) independent of the duration and spectrum of the emission in the comoving frame. These features can be tested, once light-curve modulations due to scintillation are accounted for. We provide the timescales and coherence bandwidths of the latter for a range of possibilities regarding the physical screens and the scintillation regime. Finally, if future highly resolved FRB light-curves are shown to have intrinsic variability extending down to $\sim \mu$s timescales, this will provide strong evidence in favor of magnetospheric models.
	\end{abstract}
	

	\section{Introduction}
	
	Fast Radio Bursts (FRBs) are a few milli-second duration, very bright, radio signals that have been detected between about $400$\,MHz and $7$\,GHz with peak flux density $\gae1$\,Jy. 
	The objects producing these bursts are typically at distances of a Gpc or more, \citep{Lorimer+07,Thornton+13,Spitler2014,Petroff2016,Bannister2017,Law+17,Chatterjee+17,Marcote2017,Tendulkar+17,Gajjar2018,Michilli+18,Farah2018,Shannon2018,Oslowski2019,Kocz2019,Bannister+19,CHIME2019,CHIME2019b,Ravi2019,Ravi2019b,Ravi+19}.
	
	A wide variety of models for FRBs and their high brightness and coherent radiation have been suggested in the last several years, \citep{Katz2014,Katz16,Lyubarsky14,Murase2016,Kumar+17,Metzger+17,Zhang17,Beloborodov17,Cordes2017,Ghisellini2018,LuKumar2018,Metzger+19,Thompson2019,Wang2019,Wadiasingh2019,KumarBosnjak2020}; for a recent review see \citep{Katz2018}. However, the origin of FRBs remains an unsolved puzzle. The nature of the underlying object has been identified to be magnetars only very recently by observing a FRB in the Galaxy in the radio and X-ray bands simultaneously \citep{CHIME2020,STARE2020,Mereghetti+20,Li2020,Pearlman2020,Ridania2020,Tavani2020}. Although this observation is a major landmark for FRB science, many emission models remain valid \citep{LKZ2020,Lyutikov2020,MBSM2020,Katz2020}. Furthermore, new puzzles have also emerged. For instance, the huge range of the repetition rate -- at one end lies the Galactic FRB and at the other cosmological FRBs such as FRB 121102 -- despite the rather modest difference in the energy release in the radio band is puzzling.
	One possibility is that this is indication of an intrinsic magnetar property that is vastly different between the source of the Galactic FRB and cosmological ones, such as the magnetic field \citep{MBSM2020} or the rotation period \citep{Beniamini+20}. Clearly, new types of observational constraints would be of great benefit towards advancing our understanding of FRBs
	
	A key observable that has yet to be fully exploited to decipher the underlying physics is the bursts' temporal variability. Due to the potential contributions due to propagation effects, separating temporal modulations due to external and internal affects is crucial.
	We investigate in this work what we can learn about the object and the radiation process from the temporal variability of FRB light-curves.
	
	\section{Physical processes and parameters that control light-curve variability}
	\label{sec:intvar}
	The light-curve of several FRBs show variations on very rapid timescales. For instance, FRB 170827 had an overall duration of $\sim 0.4$\,ms with strong variations on a timescale of $\sim 30 \mu$s \citep{Farah2018}. Another FRB, 181112, had a pulse
	with a very rapid rise $\sim 10 \mu$s, followed by a significantly shallower decay $\sim 0.15$\,ms \citep{Cho2020}.
	The data suggests that the variability timescale is intrinsic to the source, making it a diagnostic for the burst mechanism. There are several important timescales in the FRB phenomenology: (i) the pulse rise time, (ii) the pulse decay time, (ii) the overall pulse duration, and (iv) the time between pulses. Understanding these timescales should be useful for revealing the underlying FRB physics. 
	
	Let us consider a relativistic outflow moving with Lorentz factor $\gamma\gg 1$ that produces a radio burst at a distance $R$ from its launching point. If
	the comoving size of the outflow (in both the radial and transverse directions) is $\gae R/\gamma$ -- which is expected since the outflow from an average FRB cannot be confined by the magnetic field of a magnetar at $R\gae 10^9$ cm -- then a natural timescale for the variability, and the rise of the light-curve, is given by
	\begin{equation}
	t_0(R) \approx \frac{R}{2c\gamma^2}.
	\label{tv}
	\end{equation}
	$t_0$ corresponds to three separate and important timescales in the observer's frame. It is the observed time separation between (i) two photons
	emitted at the front of the expanding outflow at times separated by $R/c$ in the magnetar rest frame (or the lab frame), (ii) two photons emitted at the same time but separated by a radial distance $R/\gamma$ (corresponding to the causally connected thickness of the outflow) in the comoving frame and (iii) two photons emitted from the same radius and 
	at the same time, but with an angular separation of $1/\gamma$ between them (corresponding to the edge of the relativistic beaming cone).

	The natural timescale for the observed light-curve decline is also $\sim t_0$. However, the burst duration for FRB 181112 was longer than the rise time by a factor $\gae 10$. Similarly for FRB 170827 the overall duration was much longer than the variability timescale. These observations suggest that the light-curve profile is reflecting the activity of the central engine of FRBs (and/or the interaction of the outflow with the external environment). The radius of the front of the relativistic outflow therefore increases by a factor of at least a few during the course of the pulse which will generally affect the emission properties. Indeed, some FRB sub-pulses exhibit downward frequency drifts \citep{Gajjar2018,Hessels+19,CHIME2019}, in which the peak frequency of the spectrum decreases over time. Within the framework of the baryonic shock model for FRBs, these drifts can be directly related to the external density profile \citep{Margalit+20}. That being said, the recently discovered FRB 200428, exhibits a significant lack of emission at frequencies below $\sim 550$ MHz for the second pulse \citep{CHIME2020}, and a lack of frequency down-drift, which are not expected according to the shock model. As we show in detail in \S \ref{sec:temproscetral}, even if one were to terminate the maser emission process suddenly when the outflow is at radius $R$ and produced $550$ MHz photons, the observer will continue to get photons of lower frequencies which have been Doppler boosted a bit less than higher frequency photons due to the curved geometry of the shock front; cutting off the flux below $550$ MHz in the observer frame imposes severe limitations on the physical conditions at the emission region if the radiation is produced at large distance from the magnetar ($R\gtrsim10^{10}$cm).

	\subsection{Limits on intrinsic variability time and sharpness of spectral features}
	\label{sec:temproscetral}

	\begin{figure}
		\centering
		\includegraphics[width = .35\textwidth]{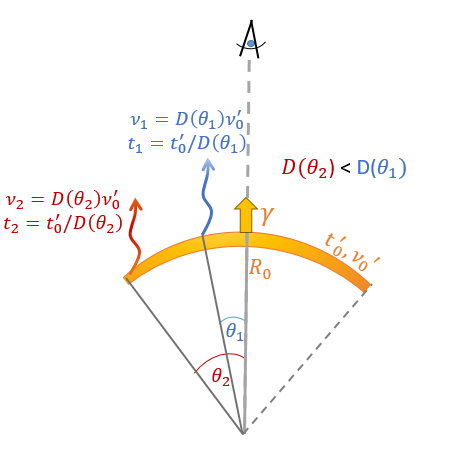}
		\caption{A shell moving with Lorentz factor $\gamma$ starts to produce coherent radio emission after crossing a radius $R_0$. The emission is assumed to be almost instantaneous in the comoving frame (occurring at a time $t_0'$) and with comoving frame spectrum that is narrowly peaked at frequency $\nu_0'$. Photons emitted along the line of sight have a Doppler factor that decreases with angle -- $D(\theta_2)<D(\theta_1)$ for $\theta_2>\theta_1$. Radiation from higher latitudes, i.e. $\theta > 0$, arrives at lower frequencies and later times.} 
		\label{fig:SchematicDoppler}
	\end{figure} 
	
	We consider a broad class of models for FRBs where a relativistic outflow from a compact object moving with Lorentz factor $\gamma$ produces coherent radio emission at a distance $R$ from the launching site (see schematic illustration in figure \ref{fig:SchematicDoppler}). We show in this section that there are constraints on the variability time scale of the light-curve, and how rapidly the specific flux can change with  frequency, for this class of models that are obtained from very general  considerations. The data should be analyzed to check for these constraints and narrow down the landscape of models for FRBs.
	
	Photons emitted isotropically in the comoving frame of the outflow moving with Lorentz factor $\gamma$ are relativistically beamed within a cone of opening angle $\sim 2/\gamma$ as seen by a lab frame observer. Thus, a lab frame observer sees a small patch of size $R/\gamma$ of a spherical source of radius $R$ that is expanding in the radial direction with a Lorentz factor $\gamma$. From the point of view of an observer, the flux of photons from those points of the outflow where the angle $\theta$  between the velocity vector and the observer line of sight is larger than $\gamma^{-1}$ -- called high latitude emission -- falls off rapidly with $\theta$. The high latitude emission limits how fast the observed flux can decline with time when the source is turned off suddenly, and it also severely restricts sharp truncations of the observed spectrum. Even in the extreme case, where the intrinsic spectrum is a delta-function in the source comoving frame peaking at a frequency $\nu_0$ in the observer frame, the declining Doppler boost from increasing latitudes ($\theta$) leads to a substantial flux at $\nu<\nu_0$ in the observer frame.
	
	To calculate these effects, consider a relativistically moving source that radiates almost monochromatic photons at frequency $\nu_0'$ in a narrow band $\delta\nu'$, at time $t_0'$, and over a short interval of time $\delta t'$; all prime ($'$) quantities are measured in the rest frame of the relativistic source which is taken to be effectively 2D, i.e. its radial width is small and variations across it are ignored. These photons will be observed at a frequency 
	\begin{equation}
	\label{eq:nuofD}
	\nu=\mathcal{D}(\theta)\nu_0'\quad{\rm where}\quad 
	\mathcal{D}(\theta)=[\gamma(1-\beta \cos\theta)]^{-1},
	\end{equation}
	is the Doppler factor, and $\theta$ is the angular position of a point on the source wrt the line joining the observer and the center of the compact object that produced the relativistic outflow (as depicted in figure \ref{fig:SchematicDoppler}); we assume that the emission is being produced well outside the light cylinder so that the velocity vector at every point in the source is in the radial  direction. The photon arrival time at the observer is related	to the comoving frame time by a Doppler shift factor:
	\begin{equation}
	t = t'/ \mathcal{D}.
	\end{equation}
	Photons emitted within a time interval $\delta t'$ from a circular ring that extends from $\theta$ to $\theta + \delta\theta$ will arrive within a time interval $\delta t$ and a frequency band $\delta\nu$ in the observer frame, which are given by
	\begin{equation}
	\delta t = t_0'\, \gamma\beta \, \theta\delta\theta + \delta t'/\mathcal{D} \quad{\rm and}\quad  \delta\nu =\mathcal{D}(\delta\nu' + \nu_0'\gamma\mathcal{D}\,\theta\delta\theta),
	\label{dtdnu}
	\end{equation}
	where we expanded the Doppler factor in terms of the small angle $\theta$ so that
	\begin{equation}
	\label{eq:D}
	\mathcal{D} \approx {2\gamma\over 1 + (\theta\gamma)^2}.
	\end{equation}
	
	The observed flux is given by
	\begin{equation}
	f(t,\nu) = \int d\Omega_{obs}\,\cos\theta_{obs}\, I_\nu  
	\quad{\rm where}\quad \theta_{obs} = \theta\, R/d,
	\end{equation}
	$I_\nu$ is the specific intensity in the observer frame, $R$ is the distance from 
	the compact object where the relativistic outflow produces the coherent 
	radiation, and $d$ is the angular-diameter distance to the object from us. 
	Making use of the Lorentz invariance of $I_\nu/\nu^3$ and azimuthal symmetry, we arrive at
	\begin{equation}
	f(t,\nu) = 2\pi  \frac{R^2\,I'_{\nu'}}{d^2} \mathcal{D}^3\, \theta \delta\theta.
	\end{equation}
	The value of $\theta\delta \theta$ depends on the extent of the ring from which photons can contribute to the observed $\nu$ at time $t$. There are two physically distinct situations: either $\delta t'/t_0'>\delta \nu'/\nu_0'$ (intrinsically narrow 
	band) or $\delta t'/t_0'<\delta \nu'/\nu_0'$ (intrinsically wide band).
	
	Let us consider first the spectrum in the comoving frame to be intrinsically narrow. In this situation, we can then take the angular width of the ring to be sufficiently small so that the time interval over which photons arrive at the observer, $\delta t$, given by eq. \ref{dtdnu}, is due to the emission time interval $\delta t'$. The maximum extent of the ring is then governed by the requirement that photons emitted across the ring, within the narrow band, have the same frequency in the observer frame, or $\gamma \mathcal{D}\,\theta \delta \theta =\delta \nu'/\nu_0'$.
	In this case, the expression for the observed flux reduces to
	\begin{equation}
	\label{eq:narrowspectrum}
	f(t,\nu) \approx \left[ {2\pi R^2 I'_{\nu'}  \delta\nu'\over \gamma d^2 \nu_0'}
	\right] \mathcal{D}^2 = f(t_{0},\nu_0) {t^2_{0} \over t^2}= f(t_{0},\nu_0) {\nu^2 \over \nu_0^2},
	\end{equation}
	where 
	\begin{equation}
	\label{eq:nunu0}
	\nu = \nu_0 \left( {t_{0}\over t} \right), \quad t_{0} \equiv 
	{t'_0 \over 2\gamma}, \quad \nu_0 \equiv 2 \gamma \nu_0',
	\end{equation}
	and $f(t_{0},\nu_0)$ is the flux received from $\theta\!=\!0$ at time $t_{0}$ and frequency $\nu_0$.

	One of the key results of this calculation is as follows. Let us consider a radio telescope that detects flux $f_1$ from a FRB at frequency $\nu_1$ at time $t_1$. The flux from the same FRB at a lower frequency $\nu_2$  and at a slightly later time $t_2=t_1 (\nu_1/\nu_2$) should be at least $f_1 (\nu_2/\nu_1)^2$. This result applies so long as the coherent radiation is produced under optically thin conditions in a relativistic outflow outside the light cylinder of the NS so that $t_1 \sim R/2c\gamma^2$. This result is derived under the highly conservative conditions which assume that 
	the radiation mechanism is turned off suddenly at time $t_1$ and the spectrum in the comoving frame of the source is infinitesimally narrow. A violation of these conditions can only increase the observed flux at $\nu_2$. 
	
	We consider next the case where the comoving spectrum is broad band $\delta \nu'/\nu_0'>\delta t'/t_0'$. Here, the observed bandwidth $\delta \nu$ can be taken to be primarily due to $\delta \nu'$, and the extent of the ring is set by $\delta t'$ such that $\gamma \beta D\theta \delta \theta =\delta t'/t_0'$. The flux is then
	\begin{equation}
	f(t,\nu)\approx \bigg[\frac{2\pi R^2 I'_{\nu'} \delta t'}{d^2 \gamma \beta t_0'}\bigg]\mathcal{D}^2
	\end{equation}
	Since the spectrum is assumed now to be broad band, we can take as an example $I'_{\nu'}\propto \nu'^{-\tilde{\beta}}$. This leads to $f(t,\nu)\propto \mathcal{D}^{2+\tilde{\beta}}$ which is the well known result for high latitude emission \citep{KP2000}. If we consider a frequency $\nu_2$ such that $\nu_2<\nu_1$ and we observe at a delayed time $t_2=t_1(\nu_1/\nu_2)$ (as above), then we find
	\begin{equation}
	\frac{f(t_2,\nu_2)}{f(t_1,\nu_1)}=\bigg(\frac{\nu_2}{\nu_1}\bigg)^{-\tilde{\beta}}\bigg(\frac{t_2}{t_1}\bigg)^{-2-\tilde{\beta}}=\bigg(\frac{\nu_2}{\nu_1}\bigg)^2=\bigg(\frac{t_1}{t_2}\bigg)^2
	\end{equation}
	exactly as in the previous case. 
	
	A corollary of this result is that if we see the flux from a FRB drop to zero sharply below a given frequency, and if we can rule out that this cutoff is due to scintillation, then that tells us that the FRB radiation is not being produced in a relativistic outflow outside the light cylinder unless the angular size of the outflow at $R$ is smaller than $\gamma^{-1}$ as viewed from the NS. 
	
	Another implication is that for a sufficiently broad-band detector, the FRB flux cannot drop off faster than $1/t^2$ if the radiation mechanism is operating in a relativistic outflow outside the light cylinder. We emphasize that these two key results are purely geometric in origin, and completely independent of the details of the radiation mechanism as long as the process involves a relativistic outflow of angular size not smaller than $\gamma^{-1}$. Thus, the flux estimate we have provided at $\nu_2$ is an absolute minimum. Observations of FRBs over a sufficiently broad frequency band, $\delta\nu/\nu\sim 1$, should be able to constrain the radiation mechanism using these results.
	
	As a specific example, we have already pointed out the case of the Galactic 
	FRB 200428 in \S \ref{sec:intvar}, in which the emission appears to cutoff abruptly below $\sim 550$MHz for the second pulse. This sharp cutoff of the spectrum potentially rules out those models which invoke radiation at $r \gtrsim 10^{10}$cm so that the pulse duration is of order $R/(2c\gamma^2)$ in the observer frame. We briefly mention two other FRB observations, which also appear to be in conflict with the basic expectations of radiation produced in a relativistic outflow. First, three bursts from FRB 121102 were detected by VLA at 2.5-3.5 GHz but not by Arecibo at 1.15-1.73 GHz, despite Arecibo being more sensitive than VLA by a factor $\sim 5$ \citep{Law+17}. Given that the Arecibo band is a factor $\sim 2$ smaller than the VLA, the expected flux from a relativistic outflow in the Arecibo band should have been at most $\sim 4$ times less than the VLA flux if the angular size of the outflow is larger than $\gamma^{-1}$. For the same FRB, \cite{Majid2020} detected a burst at 2.25 GHz but not at 8.36 GHz despite the observations being simultaneous.
	This may be consistent with the considerations above if the spectrum cuts-off above $\sim 3$ GHz. However, this also requires that the frequency drift with time is very modest, which is not trivially achieved, and may require significant fine tuning of parameters in the model.
	
\begin{figure}
		\centering
		\includegraphics[width = .4\textwidth]{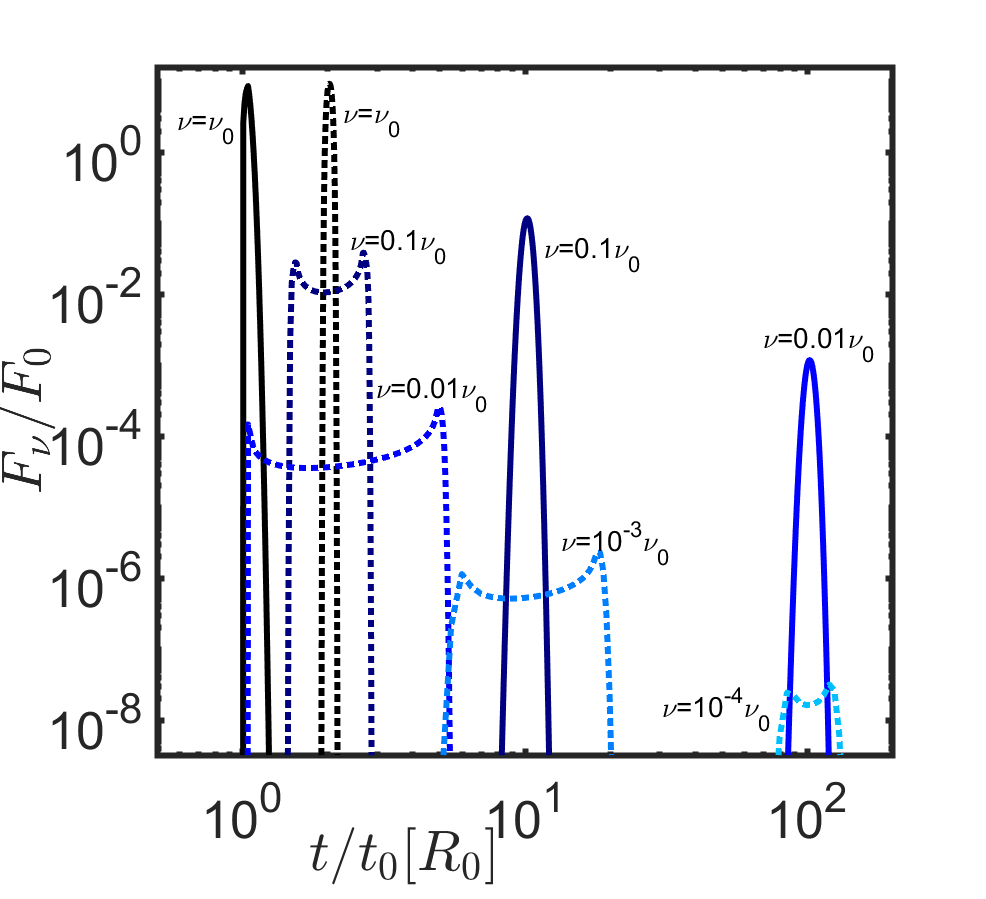}
		\caption{Pulse light-curves from a relativistic outflow moving with a Lorentz factor $\gamma$ and emitting a monochromatic spectrum (i.e. $\delta \nu'/\nu_0'<\delta t'/t_0'$) between $R_0$ and $R_0+\Delta R$ with $\Delta R/R_0=0.05$. The emission is either isotropic in the comoving frame ($\gamma'=1$; solid) or anisotropic, due to relativistic beaming with a Lorentz factor $\gamma'=10$ in the comoving frame (dotted). The calculation follows the description in \citealt{BG2016} (with $k,m,a=0$).  Time is measured relative to the arrival time of the first photon and flux is shown in arbitrary units. The observed frequency is taken to be $\nu=1,0.1,0.01,...\times \nu_0\equiv 2\gamma \nu_0'$ from black to light blue respectively.} 
		\label{fig:spectral}
	\end{figure}

	A few examples of light-curves calculated numerically following the method we have described are shown in figure \ref{fig:spectral}.
	
	\subsection{Photons re-encountering emitting shell}
	\label{sec:reencounter}
\begin{figure}
		\centering
		\includegraphics[width = .4\textwidth]{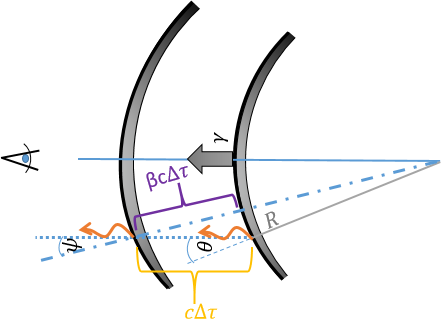}
		\caption{Schematic figure describing the geometry of a photon emitted at a radius $R$ and latitude $\theta\geq 1/\gamma$ that re-enters the emitting shell (moving with a Lorentz factor $\gamma$) after a (source frame) time $\Delta \tau$. The line of sight to the observer is towards the left of the figure.} 
		\label{fig:reencounter}
	\end{figure}

	Photons emitted along the line of sight from a latitude $\theta \geq 1/\gamma$ have a velocity component along the radial direction that is smaller than the shell that emitted them (see figure \ref{fig:reencounter}).  These photons will remain inside the shell of a finite thickness for a while and exit it at the rear end at a later time, then re-enter the shell at a larger radius and eventually escape through the front end of the shell. The frequency of the photons in the local comoving frame of the shell changes with time. One might imagine possibilities where these high latitude photons are unable to escape from the shell. For instance, escape is not possible if the photon frequency in the local comoving frame is lower than the plasma frequency \citep{Plotnikov&Sironi19}. We calculate below the local comoving frequency of the photon at the time when it re-enters the shell.
	
	To show this explicitly, we consider a photon emitted at radius $R$ and $\theta \geq 1/\gamma$, along the line of sight to the observer. For clarity, we consider first the case in which the shell is moving at a constant velocity $\beta$ (corresponding to a Lorentz factor $\gamma$). After a (source frame) time $\Delta \tau$, the photon momentum vector is at an angle $\psi$ (see figure \ref{fig:reencounter}) relative to the local shell's velocity vector (i.e. the radial direction). Equating the $x,y$ coordinates of the photon and the shell at a time $\Delta \tau>0$ we find
	\begin{equation}
	\label{eq:comparexy}
	(R+\beta c\Delta \tau)\!\cos \psi=R\!\cos \theta+c\Delta \tau \quad ; \quad (R+\beta c\Delta \tau)\! \sin \psi =R \!\sin \theta.
	\end{equation}
	With little loss of generality, we can expand these expressions in the limit of $\gamma\gg1$ and small angles. The second equality then gives an approximate expression for $\psi\approx \theta R /(R+\beta c\Delta \tau)$. Plugging this into the first equality in equation \ref{eq:comparexy}, we find 
	\begin{equation}
	\Delta \tau\approx \frac{R}{c} (\gamma^2\theta^2-1) \quad ; \quad \psi\approx \frac{1}{\theta \gamma^2}.
	\end{equation}
	Finally, we consider the frequency of the photon in the local comoving frame at the moment of re-encounter, $\nu_r'$, relative to the same quantity at the time of emission, $\nu_e'$. This is done by use of a double Doppler boost (from the local comoving frame at time of emission to the observer frame and then from that frame to the local frame at the time of the re-encounter),
	\begin{equation}
	\label{eq:comparenu}
	\frac{\nu_r'}{\nu_e'}=\frac{1-\beta \cos \psi}{1-\beta \cos \theta}\approx \frac{1+\gamma^2 \psi ^2}{1+\gamma^2 \theta^2}\approx \frac{1+(\gamma \theta)^{-2}}{1+(\gamma\theta)^2}<1.
	\end{equation}
	Equation \ref{eq:comparenu} shows that for a constant shell velocity $\frac{\nu_r'}{\nu_e'}<1$. At large latitudes $\nu_r'/\nu_e'\sim (\gamma \theta)^{-2}$.
	
\begin{figure}
		\centering
		\includegraphics[width = .45\textwidth]{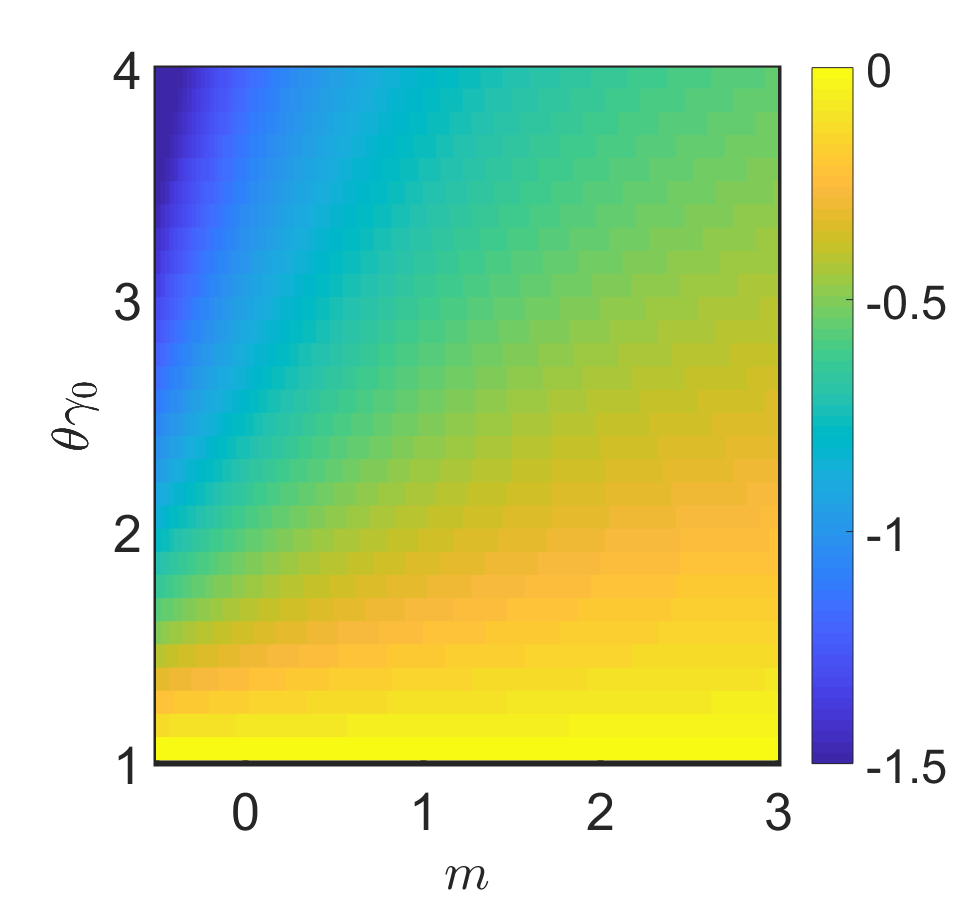}
		\caption{Change in the comoving frequency of photons [$\log_{10}(\nu_r'/\nu_e')$] that are emitted at $\theta>1/\gamma_0$ and when they re-encounter the shell at a later time. Results are shown for different emission angles and different degrees of deceleration / acceleration of the shell that is characterized by the parameter $m$ defined as $\gamma\propto R^{-m/2}$). }
		\label{fig:nuchange}
	\end{figure}

	Before discussing the implications of this result, we consider first the more general case of a shell that is decelerating (or accelerating) with radius. We consider $\gamma$ to be a power-law function of $R$: $\gamma^2\propto R^{-m}$. We denote by $\gamma_0,R_0$ ($\gamma_{\rm f},R_{\rm f}=R_0 (\gamma_f/\gamma_0)^{-2/m}$) the Lorentz factor and radius of the shell when the photon is emitted from (re-encountered by) the shell. For this more general case, equation \ref{eq:comparexy} is re-written as
	\begin{equation}
	\label{eqn:xywithm}
	R_{\rm f} \cos \psi =R_0\cos \psi +c\Delta \tau \quad ; \quad R_{\rm f} \sin \psi = R_0 \sin \theta.
	\end{equation}
	To solve these equations we must relate $\gamma_{\rm f}$ to $\Delta \tau$. The source frame time interval, $\Delta \tau$, can be related to the observer frame time interval, $\Delta t\equiv t(R_{\rm f})-t(R_0)$ via $\Delta t=\Delta \tau -\Delta R/c$. 
	Using $t\propto R/\gamma^2$ and $\gamma\propto R^{-m/2}$ we find $\gamma=\gamma_0(t/t_0)^{-m/2(m+1)}$ and hence $\Delta t= t(R_0) [(\gamma_{\rm f}/\gamma_0)^{-2(m+1)/m}-1]$. Plugging this back into equation \ref{eqn:xywithm} we find an implicit equation for $\gamma_{\rm f}$:
	\begin{equation}
	x^{m+1}-1=\xi_0-\frac{\xi_0}{x} \mbox{ for } x\equiv \bigg(\frac{\gamma_{\rm f}}{\gamma_0}\bigg)^{-2/m}\& \quad  \xi_0=(\gamma_0 \theta)^2.
	\end{equation}
	Substituting this back into equation \ref{eqn:xywithm} we can calculate $\Delta t, \psi$ and finally the frequency change
	\begin{equation}
	\frac{\nu_r'}{\nu_e'}=\frac{\gamma_{\rm f}}{\gamma_0}\frac{1-\beta_{\rm f} \cos \psi}{1-\beta_0 \cos \theta}\approx \frac{\gamma_0}{\gamma_{\rm f}} \frac{1+\gamma_{\rm f}^2 \psi^2}{1+\gamma_0^2 \theta^2}<1.
	\end{equation}
	Similar to the constant velocity case, the result in the more general case where the shell velocity changes with radius is that the photon frequency in the comoving frame is lower when the photon re-enters the shell. The values of $\nu_r'/\nu_e'$ obtained for different $\theta, m$ are shown in figure \ref{fig:nuchange}.
	
	The implication of this result is that if the peak frequency is close to the plasma frequency in the comoving frame, then photons originating from sufficiently large $\theta\gamma_0$ will have frequency below the plasma frequency when they re-enter the shell, and they will not be able to travel through the shell and reach the observer. At the same time, the effect described in this section can only constrain photons emitted at $\theta>1/\gamma_0$. Since the Doppler factor changes by a factor of $2$ between $\theta=0$ and $\theta=1/\gamma_0$, there will still be a minimum span of in the observed frequency by a factor of $2$ resulting from photons coming from those different latitudes. The spectral width of the flux in the observer frame should therefore still be at least of order unity even when the spectrum in the shell comoving frame is intrinsically very narrow and centered close to the plasma frequency.

	\subsection{Alternative scenarios leading to rapid variability}
	\label{sec:shortervar}
	We discuss a few possible deviations from the picture above that can allow for variability on a timescale shorter than that given by equation \ref{tv}. We consider their potential viability for explaining the observed variability of FRB light-curves. As we show below, the efficiency of converting relativistic outflow energy ($E_{FRB,tot}$) to FRB radiation ($E_{FRB,obs}$), $\varepsilon\equiv E_{FRB,tot}/E_{FRB,obs}$, generally decreases whenever a mechanism is introduced that reduces the variability time of the FRB lightcurve to a value smaller than that given by equation \ref{tv}. $\varepsilon$ can be considered as a product of different efficiency factors, such as the efficiency of converting blast wave energy to heat, the efficiency of the radiation mechanism (fraction of dissipated energy converted to radiation), a $k$-correction factor (fraction of emitted energy leading to the signal in the observed frequency band), etc. $\varepsilon$ being the product of these efficiency factors is smaller than the smallest factor. We stress that the considerations in this work are largely independent of the radiation mechanism. The synchrotron maser mechanism, for example, is extremely inefficient, with $\varepsilon_{\rm rad}\approx 10^{-3}-10^{-2}$ for moderate magnetization \citep{Plotnikov&Sironi19}. As a result the overall efficiency, $\varepsilon$ will be further reduced by this factor in addition to the other inefficiencies described below.
	
	\subsubsection{Small emitting clumps}
	\label{sec:clumps}
	The rise, or the variability, timescale can be smaller than $R/(2c\gamma^2)$ provided that the radiation is produced in a very small patch of the outflow of comoving size $\ll R/\gamma$.
	However, in this case, as we show below, the emitting region will have a small covering factor relative to the outflow, which leads to the efficiency of converting the outflow energy to radiation being small (see also \citealt{SariPiran1997}).

	Consider a shell with radius $R$ which consists of multiple small emitting clumps, each spanning a narrow range of latitudes. 
	For an emitting clump, extending between $\theta_1$ and $\theta_2$ such that $1\gg\theta_1,\theta_2,\geq 0$. We denote the mean angle of emission and angular width of the clump as $\bar{\theta}=\frac{1}{2}|\theta_1+\theta_2|, \delta \theta = |\theta_2-\theta_1|$ respectively. In particular, we note that for arbitrary $\theta_1,\theta_2$, $\delta \theta\leq2\bar{\theta}$. 
	We also define a dimensionless number, $\zeta$, such that the lateral size of 
	a clump is $r_s=\zeta R / \gamma$ (i.e. $\zeta=\gamma \delta \theta$).
	The time difference between photons emitted simultaneously from $\theta_1$ and $\theta_2$ is then given by
	\begin{equation}
	\label{eq:deltatclump}
	\delta t\approx \frac{R\bar{\theta}\delta \theta}{c}\geq \frac{R \delta \theta ^2}{2c}=\frac{R\zeta^2}{2\gamma^2 c}
	\end{equation}
	A short time variability can be maintained if $\delta t \ll t_0$ which requires
	$\zeta\ll 1$. 
	The fastest degree of variability in this model arises when the clump lies very close to the line of sight, $\bar{\theta}\approx \delta \theta \approx \zeta \gamma^{-1}\ll \gamma^{-1}$. In this situation, the inequality in equation \ref{eq:deltatclump} becomes an equality and one obtains $\delta t/t_0\propto \zeta^2$, and using equation \ref{eq:nunu0}, $\delta \nu/\nu_0\propto \zeta^2$. However, this situation is geometrically fine tuned given that the region from which clumps could be seen by the observer is up to a latitude of $\gamma^{-1}$ (which would correspond to slower variability). In other words, only an order unity of clumps can have $\bar{\theta}\approx \zeta$ while there are expected to be $\zeta^{-2}\gg 1$ clumps with $\bar{\theta}\approx \gamma^{-1}$. One important scenario where only emission from a small region of angular size $\bar{\theta}\approx\zeta\gamma^{-1}$ is visible to the observer, is the case of a synchrotron maser from strongly magnetized relativistic shocks. In this situation, the shock front is moving away from the shocked plasma with a Lorentz factor $\sim \sqrt{\sigma}$ \citep{Petri2007}, where $\sigma$ is the upstream magnetization of the flow. The result is that the radiation in the shocked plasma co-moving frame is beamed within an angle $\theta\sim 0.7/\sqrt{\sigma}$ around the shock normal direction \citep{Babul2020}. Since beaming is in the radial direction, the situation is equivalent to the one described above, where $\zeta\sim 0.7/\sqrt{\sigma}$ (see \S \ref{sec:anisotropic} for details) \footnote{ Notice also that in the shock front frame (as opposed to downstream plasma frame) the radiation is roughly isotropic, implying that the arrival time is well described by equation \ref{tv} so long as one takes $\gamma$ to be the Lorentz factor of the shock front in that equation}. Provided that the magnetization is large, this can lead to rapid variability ($\zeta \ll1$). However, it also results in a significant decrease in the efficiency.
	The efficiency is limited by the geometric efficiency, $\varepsilon_{\rm g}$. The latter is given by the ratio of the area of the patch that is producing maser emission and the area of the shell visible to the observer, i.e. 
	\begin{equation}
	\varepsilon\leq \varepsilon_{\rm g} \approx \zeta^2 \ll 1.
	\end{equation}
	A less fine tuned possibility is that masing clumps are randomly distributed across the shell. In this case, a typical clump has $\bar{\theta}\approx \gamma^{-1}$, and plugging back to equation \ref{eq:deltatclump} we find a variability time scale of $\delta t/t_0\approx \zeta$ and correspondingly $\delta \nu/\nu_0\approx \zeta$, which are both still narrow. However, the process remains inefficient. This is because in order to maintain a high degree of variability, i.e. to avoid the situation where emission from many clumps
	overlap and broaden the pulse width, the largest number of clumps that can contribute to a pulse of duration $t_0=\delta t\, \zeta^{-1}$ is $N=\zeta^{-1}$. Comparing the total area of the masing clumps to that of the visible shell we then get $\varepsilon\leq \varepsilon_{\rm g}\approx N\zeta^2\approx \zeta \ll 1$.

	\subsubsection{Narrow range of emission radii}
	\label{sec:narrowradii}
	Consider an outflow that starts emitting when it reaches some radius $R_0$, and stops at $R_{\rm f}=R_0+\Delta R$. If $\Delta R \ll R_0$ (and taking for simplicity $\gamma$ to be roughly constant between $R_0$ and $R_{\rm f}$) this scenario can lead to a flux rising on a timescale of (see \citealt{BG2016})
	\begin{equation}
	t_{\rm r}=\frac{\Delta R}{R_0}\, t_0(R_0),
	\end{equation}
	where $t_0$ is given by equation (\ref{tv}). The flux decreases on a much longer timescale of 
	\begin{equation}
	t_{\rm d}=t_0(R_{\rm f}) = {R_{\rm f}\over 2c\gamma^2}.
	\end{equation}
	The ratio of the rise time and the decline time ratio is roughly $\Delta R/R_0$, and this can be very small if we take $\Delta R\sim 0.05-0.1 R_0$.
	A decrease in the range of radii over which the material is emitting results in a decreased dissipation efficiency, $\varepsilon_{\rm diss}$, in this scenario. The latter is given by the ratio of the comoving shell thickness that is accessible within a time $t_{\rm r}'$, and the total comoving thickness of the shell, $L'$:
	\begin{equation}
	\varepsilon\leq \varepsilon_{\rm diss}=\min\bigg(\frac{\beta'}{2}\frac{\Delta R}{\gamma L'},1\bigg)
	\end{equation}
	where depending on the energy extraction mechanism, $\beta'<1$ is either the (comoving) velocity of matter falling into the reconnection layer or the propagation speed of a shock. If we further assume that $L'>R_0/\gamma$ (since the latter is the causally connected width), we find $\varepsilon\leq \varepsilon_{\rm diss}<0.5\beta' \Delta R/R_0 \ll 1$.
	
\begin{figure}
		\centering
		\includegraphics[width = .45\textwidth]{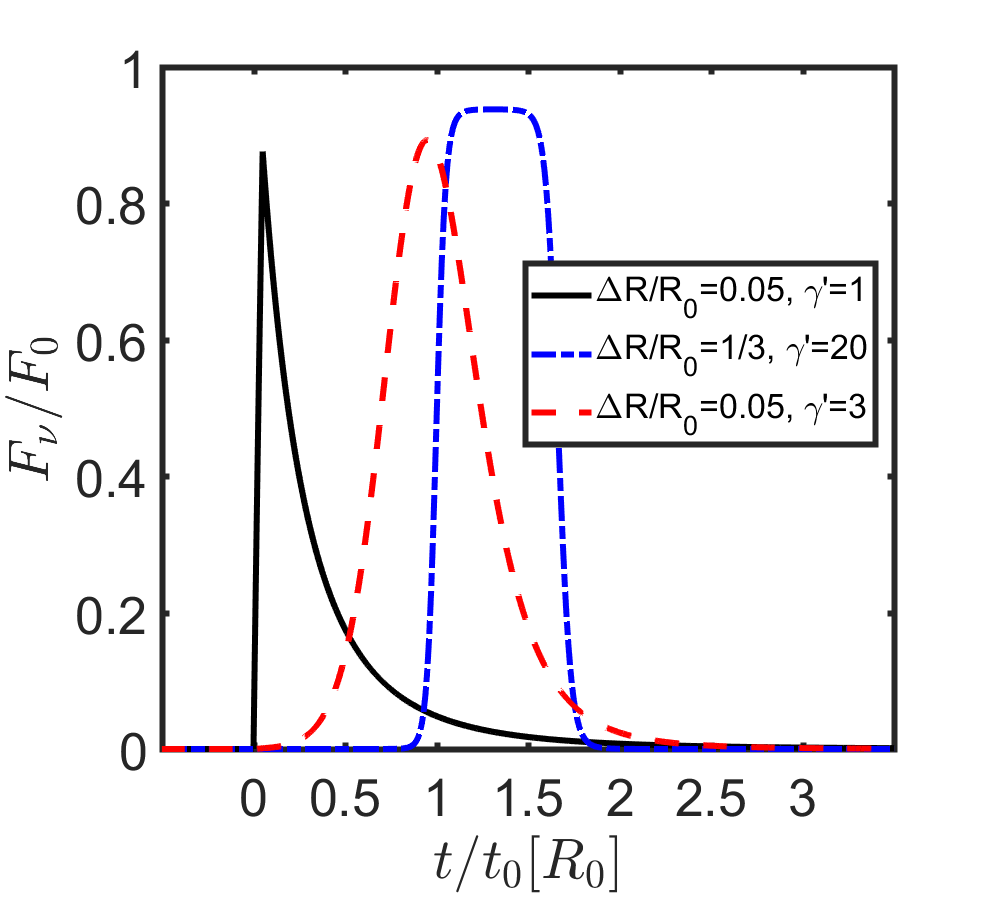}
		\includegraphics[width = .45\textwidth]{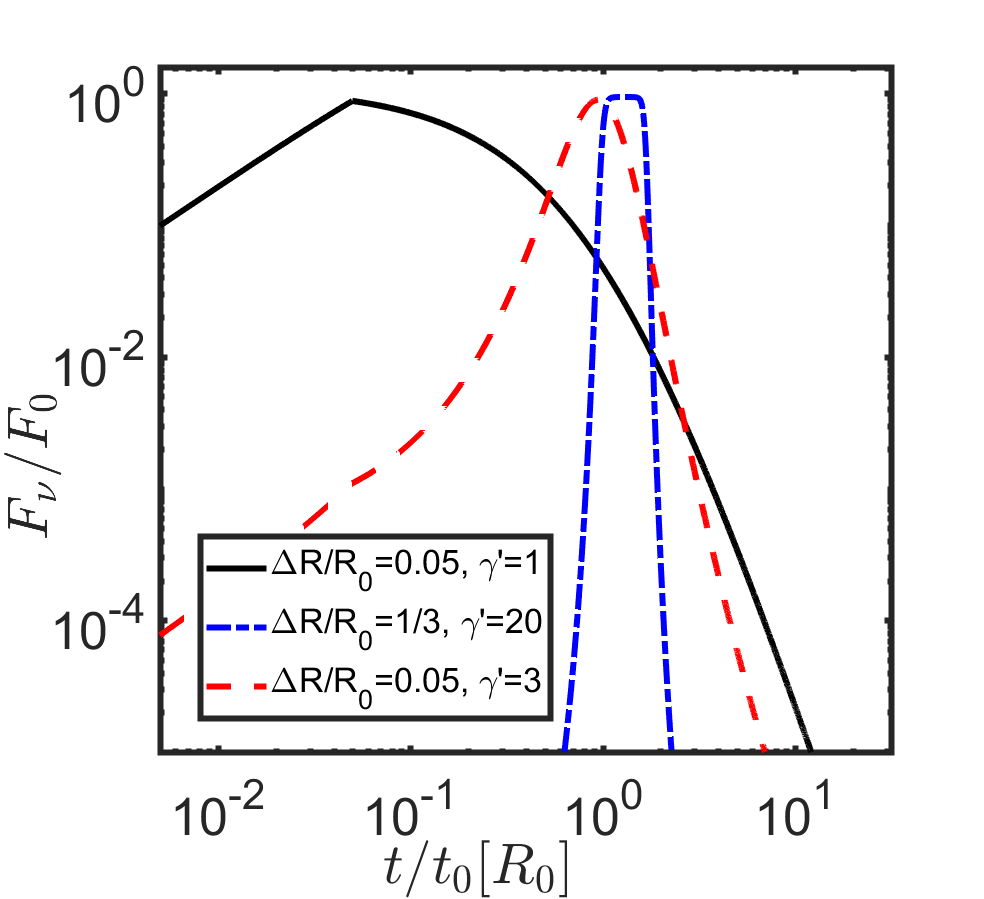}
		\caption{Pulse light-curves from the same setup as in figure \ref{fig:spectral}, but for a  narrow range of emission radii $\Delta R/R_0\!\ll \!1$ and observations at the peak frequency, $\nu_0$.}
		\label{fig:var}
	\end{figure}

	An illustration of a typical pulse light-curve that would be obtained in this case is shown as a solid black curve in figure \ref{fig:var}.
	The flux is rising for $t<t_{\rm r}$ and declining afterward. Note however, that the majority of the decline occurs in earnest only at $t\gtrsim t_{\rm d}$. The latter corresponds to the time at which the observed signal becomes dominated by photons 
	that are emitted at latitudes greater than $1/\gamma$ relative to the line of sight to the observer (which is also when the decrease in the Doppler factor becomes very rapid). This behavior is much more readily noticeable in a log-log plot, which could be used in future observations to test such a scenario. More generally, since the model described in this subsection is geometrical in nature, its predictions are robust and a reasonably good representation of the light-curve can be obtained without needing to `put in by hand' any unknown temporal activity by the central engine. Note also that in this scenario the FRB spectrum does not significantly change during the observed pulse, as the radius of the outflow
	does not evolve significantly during the emission process.
	
	Although attractive for explaining the light-curve of FRB 181112, an obvious drawback of this scenario is that it does not provide a satisfying explanation for the light-curve of those FRBs which feature `sub-pulses' that exhibit comparable rise and decay time-scales, which are both much shorter than the overall pulse width. We also emphasize that in this model, the engine must still be varying on a rapid timescale of the order of $0.1-1$ ms, or the observed time difference between consecutive pulses.
	
	\subsubsection{Radial evolution of the spectral peak}
	\label{sec:freqradius}
	Here we imagine a similar setup to the one explored in \S \ref{sec:narrowradii}, but we allow for the peak frequency to change with radius, such that
	\begin{equation}
	\nu_p'=\nu_0'\bigg(\frac{R}{R_0}\bigg)^{\delta}
	\end{equation}
	and (in order to have an appreciable evolution of the peak frequency) we relax the assumption regarding a narrow range of emitting radii, adopted above. We assume that the source is coasting at a constant speed with Lorentz factor $\gamma$, and maser emission is suddenly turned on when the shell is at radius $R_0$; note however that the results regarding the rise time presented below persist even if the emissivity changes only gradually with radius. Since $\gamma$ is time independent, the observer frame time is directly related to the radius $t \approx R/(2c\gamma^2) \propto R$.
	
	The observed frequency relates to the peak frequency as
	\begin{equation}
	\label{eq:specR}
	\frac{\nu}{\nu_p}=\bigg(\frac{\nu}{\nu_0}\bigg)\bigg(\frac{t}{t_0(R_0)}\bigg)^{-\delta}
	\end{equation}
	where $\nu_0=2\gamma \nu_0'$.
	If the emitted spectrum is intrinsically narrow in the comoving frame, this can lead to a sharp peak in the observed band, as we show below. It is important to stress however that the observer frame spectrum will not be narrow (it will be subject to the same constraints presented in \S \ref{sec:temproscetral} and will be broadened even further due to the evolution of the peak frequency in the comoving frame).
	
\begin{figure}
		\centering
		\includegraphics[width = .4\textwidth]{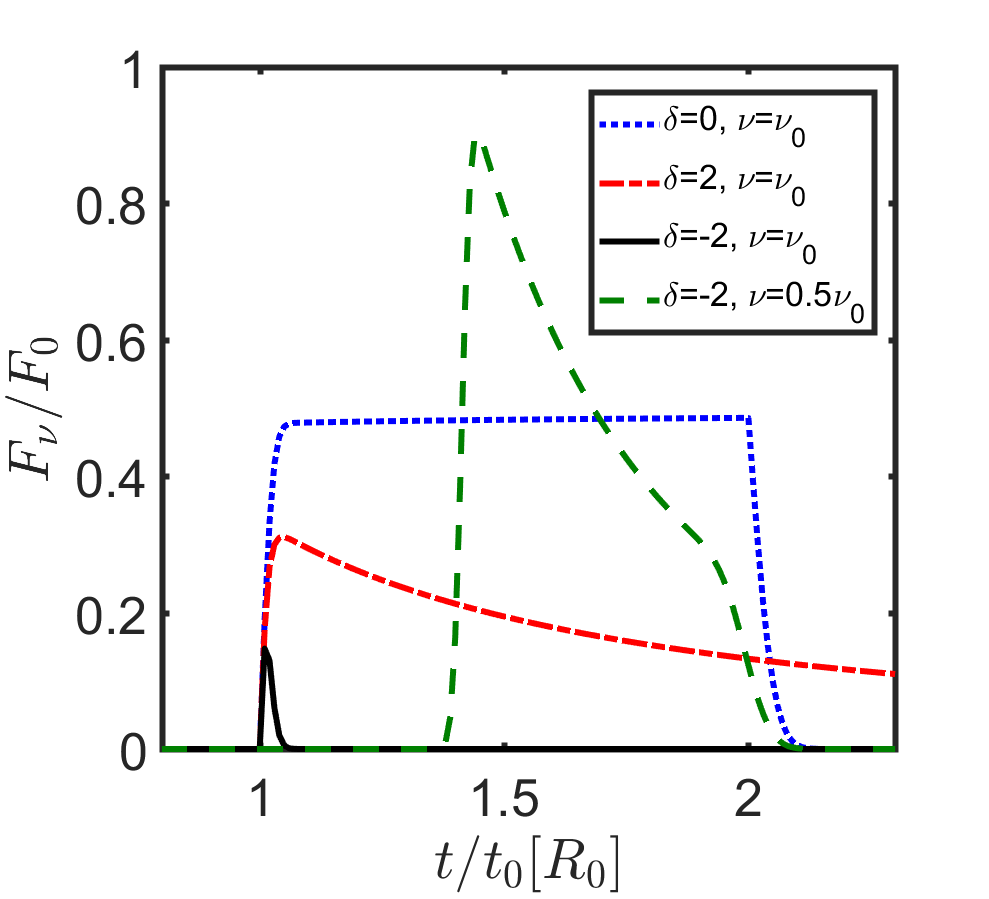}
		\caption{Pulse light-curves from the same setup as in figure \ref{fig:spectral}, but allowing for the peak frequency to evolve with $R$ as $\nu_p'\propto R^{\delta}$ and given a constant emissivity between $R_0$ and $R_{\rm f}=2R_0$. The dotted curve corresponds to the situation with no radial evolution of the peak frequency ($\delta=0$). The dot-dashed (solid) curve depicts a pulse with similar properties except for $\delta=2$ ($\delta=-2$), implying that the peak at $t>t_0[R_0]$ becomes progressively greater (smaller) than the observed frequency. These correspond to case (ii) in \S \ref{sec:freqradius}. Finally, the dot-dashed line shows the dependency on the observed frequency, by illustrating the observed signal for a similar pulse as in the solid line, observed at half the frequency.}
		\label{fig:nuofR}
	\end{figure}

	There are three relevant cases of interest to consider in this case. For clarity, we initially differentiate between the cases using the limit $R_{\rm f}\gg R_0$ and then comment, when discussing case (i), on the changes when $R_{\rm f}/R_0$ is finite.
	\begin{enumerate}
		\item ($\nu<\nu_0$ and $\delta<0$) or ($\nu>\nu_0$ and $\delta>0$). In this case the peak frequency sweeps through the observed band. From equation \ref{eq:specR}, we obtain the time of the observed peak when $\nu=\nu_p$:
		\begin{equation}
		t_{\rm p}=\bigg(\frac{\nu}{\nu_0}\bigg)^{1/\delta}t_0(R_0)
		\end{equation}
		From this expression we see that for a finite $R_{\rm f}/R_0$ the peak frequency sweeps through the observing band only when $(\nu/\nu_0)^{1/\delta}<R_{\rm f}/R_0$. If this condition is not satisfied then the appropriate physical regime is (ii) if $\nu<\nu_0$\footnote{Note however, that although the qualitative behavior is as described in (ii), the specific expressions for the peak and rise times are different in this situation than presented in case (ii) due to the fact that the high latitude emission will arrive from $R_{\rm f}$ rather than $R_0$.} or (iii) if $\nu>\nu_0$.
		
		Another relevant timescale, $t_*$, is that corresponding to the arrival time of photons that were Doppler boosted from the peak of the spectrum in the
		source comoving frame to $\nu$ in the observer frame; $t_*$ is given by
		\begin{eqnarray}
		&t_*=\left\{ \begin{array}{l}t_0(R_0)(\nu_0/\nu)\quad \mbox{ for }\delta<0\\
		t_0(R_0)(\nu_0/\nu)(R_{\rm f}/R_0)^{1+\delta} \quad \mbox{ for }\delta>0,
		\end{array} \right.
		\end{eqnarray}
		In particular $t_*<t_{\rm p}$ for $-1<\delta<0$, and $t_*>t_{\rm p}$ otherwise. The flux in between $\min(t_{\rm p},t_*)$ and $\max(t_{\rm p},t_*)$ evolves as a power-law. To see this, We can relate the angles and radii being Doppler boosted to the given observed frequency through $\nu=\mathcal{D}(\theta) \nu_0' (R/R_0)^{\delta}$. Plugging into the relation $t(R)=t_0'(R)/\mathcal{D}(\theta)$ we find $\mathcal{D}\propto t^{-{\delta\over \delta+1}}$ which, assuming a constant peak emissivity as a function of radius ($L'_{\nu_p'}=const$), leads to\footnote{This is a rising flux when $-1<\delta<0$ which is equivalent to $t_*<t_{\rm p}$. Therefore $t_{\rm p}$ is always the peak of the observed light-curve.} $f\propto \mathcal{D}^2\propto t^{-{2\delta \over \delta+1}}$.

		If the spectrum is intrinsically narrow, with $\delta \nu'\ll \nu_0'$, the result will be a rapidly rising light-curve. The observed rise time is given by the difference between $\nu_0'$ and $\nu_0'-\delta \nu'$ passing through the observed band:
		\begin{eqnarray}
		&t_{\rm r}=\left\{ \begin{array}{l}\frac{\delta \nu'}{\nu_0'}\bigg(\frac{\nu}{\nu_0}\bigg)t_0(R_0) \quad \mbox{ for }-1<\delta<0\\
		\bigg|\frac{1}{\delta}\bigg|\frac{\delta \nu'}{\nu_0'}\bigg(\frac{\nu}{\nu_0}\bigg)^{1/\delta}t_0(R_0) \quad \mbox{ else},
		\end{array} \right.
		\end{eqnarray}
		The sharp rise of the spectrum, comes at the price of a reduced `K-correction efficiency', $\varepsilon_{\rm K}$ (defined here as the fraction of emitted photon energy resulting as a signal in the observed band). This is because as $\nu$ is driven further from $\nu_0$ there is an increasing fraction of the emitted radiation that will be missed by the observer.
		The expression for the K-correction efficiency can be straightforwardly derived for the three different sub-regimes ($\delta\leq-1$, $-1<\delta<0$, $\delta\geq0$) by integrating the light-curve to compute the total energy received in the observed frequency band and comparing that to the total bolometric energy emitted by the source. 
		As an illustration, for $\delta<-1$, the result is
		\begin{eqnarray}
		&\varepsilon\leq\varepsilon_{\rm K}\approx\frac{ 2 (\delta+1)t_{\rm p} \nu [(t_*/t_{\rm p})^{1-\delta\over 1+\delta}-1]}{(\delta-1)t_0(R_0)\nu_0[(R_{\rm f}/R_0)^{\delta+1}-1]} 
		\label{eq:efficiencynusweep}
		\end{eqnarray}
		where the factor $[(R_{\rm f}/R_0)^{\delta+1}-1]/(\delta+1)$ arises due to the fact that the bolometric luminosity $L\propto \nu_p L_{\nu_p}\propto R^{\delta}$ and similarly the factor in the numerator is due to integration over the observed  flux $f\propto t^{-{2\delta\over \delta+1}}$. The efficiency implied by equation \ref{eq:efficiencynusweep} is generally very small. This can be more easily seen in the limit $R_{\rm f}\gg R_0, t_*\gg t_{\rm p}$, when the K-correction efficiency reduces to 
		\begin{equation}
		\varepsilon\leq \varepsilon_{\rm K}\leq\frac{(\delta+1)}{2(\delta-1)} \bigg(\frac{\nu}{\nu_0}\bigg)^{\delta+1\over \delta}\ll 1.
		\end{equation}
		\item $\nu<\nu_0$ and $\delta>0$. In this case the observed frequency along the line of sight always remains below $\nu_p$. This is similar to the standard set-up explored in \S \ref{sec:temproscetral}, \S \ref{sec:narrowradii}, (with the addition that the peak frequency is moving further away from the observer band over time). Since emission from along the line of sight peaks above the observed band, the received emission is dominated by a latitude $\theta_1>0$ such that $\mathcal{D}(\theta_1)=\nu/\nu_0'$. The peak flux is obtained at a time
		\begin{equation}
		t_{\rm p}=t_0(R_0)\frac{\nu_0}{\nu}.
		\end{equation}
		The rise time is again rapid, as it corresponds to the difference in arrival times from a narrow ring spanning between $\theta_1$ and $\theta_2$ defined by $\mathcal{D}(\theta_2)=\nu/(\nu_0'-\delta \nu')$. Using equation \ref{eq:D} we find
		\begin{equation}
		t_{\rm r}=\frac{t_0'[R_0]}{\mathcal{D}(\theta_1)}-\frac{t_0'[R_0]}{\mathcal{D}(\theta_2)}\approx t_0(R_0) \frac{\delta \nu'}{\nu_0'} \frac{\nu_0}{\nu}.
		\end{equation}
		Since the peak of the emission from the line of sight material is missed by the observer, the K-correction efficiency in this scenario is even smaller than in case (i). It can be estimated as follows:
		\begin{eqnarray}
		\varepsilon\leq \varepsilon_{\rm K}\approx \frac{2(\delta+1) \bigg[(R_{\rm f}/R_0)^{1-\delta}-1\bigg]}{(\delta-1)\nu_0 \bigg[(R_{\rm f}/R_0)^{\delta+1}-1\bigg]} \bigg(\frac{\nu}{\nu_0}\bigg)^2\ll 1
		\end{eqnarray}
		where we have used the result that $f(t_{\rm p},\nu)/f(t_0,\nu_0)=(\nu/\nu_0)^2$ as derived in \S \ref{sec:temproscetral}.
		\item $\nu>\nu_0$ and $\delta<0$. In this case the observed band always remains above $\nu_p$. The result is that no flux is seen by the observer.
	\end{enumerate}
	Some examples of light-curves corresponding to the different cases discussed above are given in figure \ref{fig:nuofR} for a shell that is emitting a signal with an intrinsic spectrum $\delta \nu'/\nu_0'\ll 1$ while it propagates between $R_0$ and $R_{\rm f}=2R_0$.
	
	\subsubsection{Anisotropic emission in the comoving frame}
	\label{sec:anisotropic}
	Several physical scenarios for FRB emission may lead to anisotropic emission in the comoving frame of the outflow. One such case is if the dissipation is dominated by magnetic reconnection in a high $\sigma$ flow with an ordered field orientation. Under such conditions, the plasma flowing out of the reconnection region, and producing the emission, is moving with a Lorentz factor $\gamma'\gtrsim 2$ with respect to the mean rest frame of the outflow \citep{Lyubarsky2005} and its radiation becomes narrowly beamed {\it in the comoving frame}. The shape of the light-curves arising from this configuration were studied in detail by \cite{BG2016}, and we refer the reader to that paper for a more in-depth discussion of this possibility.

	An attractive feature of this scenario, is that the
	rise time of the light-curve can be significantly shorter than $t_0$ (by either $\gamma'^{-1}$ or $\Delta R/R$, depending on the value of both parameters as well as on how the emissivity evolves with radius; see table 1 and figure 7 of \cite{BG2016} for a comprehensive coverage of the parameter space).
	This scenario shares the advantages of the narrow range of emitting radii (\S \ref{sec:narrowradii}), with regards to the implications for the cyclotron maser mechanism and the lack of spectral evolution during a pulse. In addition, it can account for both symmetric and asymmetric pulses (see figure \ref{fig:var}). It also allows for a high latitude emission decline that is steeper than the $f_{\nu}\propto t^{-2-\tilde{\beta}}$ described above.
	The final notable feature of this scenario, is that it leads to an earlier peak of the emission at $\nu\ll \nu_0$ (but with the same $(\nu/\nu_0)^2$ suppression as discussed in \S \ref{sec:temproscetral}). We show this below, for a specific geometry of the emitters in the comoving frame.

	Consider first the situation in which the emitters are moving purely in the radial direction. If in addition their distribution is uniform across the emitting surface of the jet, then the situation is equivalent to isotropic emission in the comoving frame with a modified value of the bulk Lorentz factor $\gamma$ and rapid variability will be difficult to achieve (see \S \ref{sec:intvar}). If alternatively the distribution of such emitters across the jet surface is patchy, then rapid variability becomes possible, but at a price of a significantly reduced efficiency (see \S \ref{sec:clumps}). Therefore, it is the perpendicular components of the emitters' motion that are worth exploring in more detail.
	For concreteness we assume that the emitting plasma is moving with a Lorentz factor $\gamma'$ relative to the bulk outflow in two opposite directions ($\hat{\beta}'=\pm \hat{x}'$) which are perpendicular to the radial coordinate, as in \cite{BG2016}. Rewriting equations 14, B4 of \cite{BG2016} with $m=0$ (Lorentz factor constant with radius) and $x=1$ (since we are considering monochromatic emission) we find the following relation
	\begin{equation}
	\label{eq:xi}
	(\xi+1)^{-1}=\frac{\nu}{\nu_0}\gamma'^2 \bigg(1\pm \beta' \frac{2\xi^{1/2}}{1+\xi}\bigg) \quad ; \quad \xi=(\gamma \theta)^2
	\end{equation}
	where the $\pm$ accounts for the different directions of motion in the comoving frame.
	For $\gamma'\gg1,\nu\ll \nu_0$ equation \ref{eq:xi} has two solutions. Since $t/t_0= 1+\xi$ this leads to a double peaked signal (for $\nu_0\geq\nu\gamma'^2$ these correspond to the different Doppler boosting from the material moving in opposite directions in the bulk frame). Since the emission from the `counter mini-jet' can only be seen for $\nu\leq\nu_0\gamma'^{-2}$, the result is that the anisotropic emission leads to less of a delay for moderate $\nu/\nu_0$ (but still with a suppression that scales approximately as $\nu^{-2}$). Once $\nu\ll\nu_0$ the spectral suppression, time delay and pulse broadening approach the same scalings found in equation \ref{eq:narrowspectrum} for the isotropic emission.
	
	\section{Extrinsic variability}
	\label{sec:Extvar}
	Radio waves from a transient source passing through a medium with fluctuating density can smear out small scale intrinsic temporal fluctuation of the source, and possibly impose fluctuation on a longer timescale. We analyze the timescales for these effects. We discard redshift factors in all derivations and focus instead on the physics of fluctuations, in particular the viscous dissipation and the size of smallest eddies in the inertial subrange for the Kolmogoroff spectrum of turbulence, and its impact on FRB light-curves. Appropriate redshift factors can be found in many published works, e.g. \cite{MacquartKoay13,Xu2016}.
	
	The phase change suffered by EM waves of frequency $\omega$ passing through 
	a turbulent eddy of size $\ell$, in an ionized medium, is
	\begin{equation}
	\delta\phi \sim (k \ell) \omega_p^2/(2\omega^2) \sim {q^2 \ell \lambda 
		\delta n_e\over m c^2},
	\end{equation}
	where $k = \omega/c$, $\omega_p$ is plasma frequency, $\delta n_e$ is the
	electron density fluctuation associated with eddies of size $\ell$, and 
	$\lambda = 2\pi/k$. We consider a powerlaw density fluctuation
	in the inertial subrange between length scale $\ell_{\rm min}$ and $\ell_{\rm max}$
	given by
	\begin{equation}
	\delta n_e(\ell) = n_e (\ell/\ell_{\rm max})^{\alpha}.
	\end{equation}
	The index $\alpha=1/3$ for Kolmogoroff density fluctuations. The largest 
	eddy size, $\ell_{\rm max}$, is the scale at which energy is injected to maintain
	the turbulence, and the smallest scale $\ell_{\rm min}$ is determined by dissipation
	physics of turbulence. The cumulative 
	phase change for a wave moving through a turbulent medium of thickness 
	$L$ is
	\begin{equation}
	\Delta\phi(\ell) \sim (L/\ell)^{1/2}\delta\phi \sim {q^2 n_e \lambda 
		L^{1\over2}\ell^{2\alpha+1\over2}\over m c^2 \ell_{\rm max}^{^{\alpha}} }.
	\label{delphi}
	\end{equation}
	The strong scattering case, $\Delta\phi\gae 1$, is dominated by eddies of size\footnote{The contribution to the phase change of waves passing through the scattering screen decreases almost linearly with decreasing eddy size. So, although eddies smaller than $\ell_\pi$ scatter 
		waves by a larger angle -- which scales as $\ell^{-(1-2\alpha)/2}$, see 
		equation \ref{deltheta} -- because 
		the cumulative phase change due to the smaller eddies over the entire thickness
		of the screen is less than one radian, they contribute little to modulating 
		the flux of radio waves traveling through the medium. Eddies that scatter 
		the wave the most 
		{\it and} change the wave phase by $\gae 1$ radian are eddies of size
		$\ell_\pi$ and they are the most effective for diffractive scintillation.}
	\begin{equation}
	\ell_\pi \sim \left( {m c^2\over q^2 n_e\lambda}\right)^{
		{2\over2\alpha+1}} \ell_{\rm max}^{{2\alpha\over 2\alpha+1}} 
	L^{-{1\over 2\alpha+1}} .
	\label{lpi}
	\end{equation}
	This expression is valid only if $\ell_\pi>\ell_{\rm min}$, the
	minimum size for eddies in the scattering screen. 
	
	\medskip
	\noindent{\it\bf Smallest size eddies}
	\medskip
	
	Let us consider that the medium has magnetic field of strength $B$, 
	temperature $T$ and that the average thermal velocities of electrons and 
	protons are $\bar{u}_e$ and $\bar{u}_p$. The Larmor radius for a particle 
	of mass $m$ is
	\begin{equation}
	l_B \approx {(3 m k_B T)^{1/2} c\over q B },
	\label{RB}
	\end{equation}
	where $k_B$ is the Boltzmann constant.
	The smallest eddy is unlikely to be of size $<l_B$ for
	electrons, which is usually the smallest length scale for the low-density
	cosmic plasma that typically has very large collisional mean free-path.
	Furthermore, eddies should be able to survive viscous damping. 
	The minimum eddy size, $\ell_{\rm min}$, in the turbulent cascade is determined
	by equating the viscous dissipation time to the eddy turnover time. We 
	provide a rough estimate of this scale. 
	
	The mean free path of protons at temperature $T$ for the Coulomb scattering is
	\begin{equation}
	l_{\rm mf} \sim k_B^2 T^2/(n_e\, q^4\,\ln\Lambda)\sim 2\times 10^{19}\,T_4^2/n_{_{-7}}\;
	{\rm cm},
	\label{lmf}
	\end{equation}
	where $\ln\Lambda\sim 20$ is the Coulomb logarithm and where unless otherwise stated we use the convention $q_x\equiv q/10^x$ in cgs units here and elsewhere in the paper. Let us consider an eddy 
	of size $\ell$ and speed $u_\ell$. Assuming that $l_B\ll l_{\rm mf}$, the 
	probability that a proton in the eddy traveling a distance $l_B$ will 
	collide with another proton is $l_B/\ell_{\rm mf}$. The relative velocity 
	of collision between these protons is $l_B \left| (\partial u/\partial x)
	\right|\sim u_\ell (l_B/\ell)$. Therefore, the fraction of a proton's energy lost in the collision is $\sim (l_B/\ell)^2$. Since the probability of collision after traveling $l_B$ is $l_B/\ell_{\rm mf}$, the fraction of energy lost in one Larmor time, $l_B/\bar{u}_p$, is $\sim l_B^3/(\ell^2\ell_{\rm mf})$. The viscous damping time for the eddy is thus\footnote{This timescale is for dissipation of particle momentum perpendicular to the local magnetic field. The dissipation time 
		for the longitudinal component of momentum is different, which leads to eddies that are elongated along the magnetic field direction. \cite{Goldreich1995} showed that the ratio of the eddy size in the longitudinal and transverse direction is $\sim (\ell_{\rm max}/\ell)^{1/3}$.
		We are ignoring the elongated shape of eddies on wave propagation by assuming that magnetic fields are highly tangled in turbulent cascades. This, however, is not valid as the eddy size approaches the Larmor radius.}
	\begin{equation}
	t_{\rm vis} \sim \frac{l_B}{\bar{u}_p}\, \frac{l^2 l_{\rm mf}}{l_B^3}\sim 
	{ \ell^2\ell_{\rm mf} \over l_B^2 \bar{u}_p}.
	\end{equation}
	We define the viscous length scale by equating the viscous time with eddy turnover time of $\ell/u_\ell$, which yields
	\begin{equation}
	\ell_{\rm vis} \sim { l_B^{3/2} \ell_{\rm max}^{1/4} \over M_t^{3/4} 
		\ell_{\rm mf}^{3/4} } \quad{\rm where}\quad M_t = {u_{_{\ell_{\rm max}}}\over 
		\bar{u}_p},
	\label{lmin}
	\end{equation}
	is the Mach number of turbulence at scale $\ell_{\rm max}$. 
	We made use of $u_\ell = u_{\ell_{\rm max}} (\ell/\ell_{\rm max})^{1/3}$ for 
	inertial range eddies in deriving this result.
	Substituting for the Larmor radius (eq. \ref{RB}) and particle mean-free path (eq. \ref{lmf}) we arrive at
	\begin{equation}
	\ell_{\rm vis} \sim {\ell_{\rm max}^{1\over4}\over M_t^{3\over4}} \left[ 
	{3 m c^2 q^2 n_e \,\ln\Lambda\over k_B T B^2 }\right]^{3\over4} \sim 
	(10^8\, {\rm cm})\, {\ell_{\rm max,20}^{1\over4}\over M_t^{3\over4}} 
	{n_e^{3\over4}\over T_4^{3\over4} B_{-6}^{3\over2} }.
	\label{lmin1}
	\end{equation}
	The expression for $\ell_{\rm vis}$ is consistent with the result in \S5.2 
	of \cite{Goldreich1995}.
	The size of the smallest eddy is given by
	\begin{equation}
	\ell_{\rm min}=\max(\ell_{B},\ell_{\rm vis}).
	\label{lmin2}
	\end{equation} 
	
	\medskip
	\noindent{\it\bf Deflection angles and temporal variability}
	\medskip
	
	The angle by which an EM beam passing through the turbulent screen is
	deflected is
	\begin{equation}
	\delta\theta \sim {|\vec\nabla_\perp \Delta\phi|\over k} \sim \left\{
	\begin{array}{l}
	\hskip -5pt {\lambda\over \pi\ell_\pi} \hskip 122pt  \ell_\pi > \ell_{\rm min}
	\\ \\
	\hskip -5pt {\lambda\Delta\phi(\ell_{\rm min})\over \pi\ell_{\rm min}} 
	\sim {\lambda\over \pi\ell_{\rm min}} \left[{\ell_{\rm min}\over\ell_{\pi}}
	\right]^{{2\alpha+1\over2}} \hskip 10pt  \ell_\pi < \ell_{\rm min}
	\end{array}
	\right.
	\label{deltheta}
	\end{equation}
	where $\vec{\nabla}_\perp$ is derivative taken in the direction perpendicular
	to the wave propagation. The deflection angle is proportional to 
	$n_e\, \ell_{\rm min}^{(2\alpha-1)/2}L^{1/2}/\ell_{\rm max}^\alpha$ when
	$\ell_\pi < \ell_{\rm min}$; while for $\ell_\pi>\ell_{\rm min}$,  
	$\delta\theta \propto [n_e^2L/\ell_{\rm max}^{2\alpha}]^{1/(2\alpha+1)}$.
	
	Consider a source at a distance $d_{\rm so}$ from the observer. The distance
	between the source and the scattering screen is $d_{\rm sl}$, and the
	distance between the lens and the observer is $d_{\rm lo}$. If a wave packet
	from the source is scattered by an angle $\delta\theta$ by the turbulent
	screen then a straightforward geometrical calculation shows that it will 
	arrive at the observer with a delay of
	\begin{equation}
	\label{eq:deltat}
	\delta t = {(\delta\theta)^2\over 2c} {d_{\rm lo} d_{\rm sl}\over d_{\rm so}} = 
	{R_{\rm F}^2(\delta\theta)^2\over 2c\lambda},
	\end{equation}
	where
	\begin{equation}
	R_{\rm F} = \left[ {\lambda d_{\rm lo} d_{\rm sl}\over d_{\rm so}}\right]^{1/2}
	\label{Rf}
	\end{equation}
	is the Fresnel scale. Making use of equations (\ref{deltheta},\ref{eq:deltat}) we find 
	\begin{equation}
	\delta t \sim {R_{\rm F}^2\over 2\pi^2\lambda c} 
	\left\{  \begin{array}{l}
	\hskip -5pt {\lambda^2\over\ell_\pi^2} \propto 
	\lambda^{4\alpha+6\over 2\alpha+1} \hskip 79pt \ell_\pi > \ell_{\rm min} \\ \\
	\hskip -5pt {\lambda^2\over \ell_{\rm min}^2} \left[{\ell_{\rm min}\over
		\ell_\pi}\right]^{2\alpha+1} \propto \lambda^4 
	\hskip 37pt   
	\ell_\pi < \ell_{\rm min}
	\end{array}
	\right.
	\label{del_t}
	\end{equation}
	Physically, $\delta t$
	is the observed duration of a source which has delta-function pulse profile.
	So, the turbulent scattering screen smooths out intrinsic fluctuations in the
	light-curve on timescales shorter than $\delta t$.
	
	The scattering can also imprint fluctuations on the observed light-curve. 
	One of the timescales for fluctuations imposed by the scattering screen 
	is the eddy turnover time, 
	\begin{equation}
	\delta t_{\rm ed} \sim \max\left\{ \ell_\pi, \ell_{\rm min}\right\}/v_{\rm ed},
	\label{ext1}
	\end{equation}
	where $v_{\rm ed}$ is eddy speed. 
	We receive waves from an area of the {\it screen} of radius $R_{\rm scat} \sim 
	(R_{\rm F}^2/\lambda)\delta\theta$ since rays are deflected by eddies by an angle 
	$\delta\theta$ given by equation (\ref{deltheta})
	\begin{equation}
	R_{\rm scat} \sim {R_{\rm F}^2\,\delta\theta\over \lambda} 
	\sim {R_{\rm F}^2 \over \pi}
	\left\{
	\begin{array}{l}
	\hskip -5pt {1\over \ell_\pi}  \hskip 70pt  
	\ell_\pi > \ell_{\rm min}
	\\ \\
	\hskip -5pt  {1\over \ell_{\rm min}} \left[{\ell_{\rm min}\over\ell_{\pi}}
	\right]^{{2\alpha+1\over2}}
	\hskip 10pt  \ell_\pi < \ell_{\rm min}
	\end{array}
	\right.
	\label{rscat}
	\end{equation}
	Another externally imposed variability time for the observed flux is the time it takes for the turbulent screen to move a distance $\sim\ell_\pi$ in the plane of the sky, i.e. transverse to the observer-source line, so that the scintillation pattern shifts at the observer location by one fringe width. To see how this comes about, let us consider moving the screen by a distance $\delta l$ in the transverse direction while keeping the turbulent eddies frozen. In this case, the phase shift of the wave passing through a coherent patch (of size $\ell_\pi$) changes by $\delta\phi\sim 2\pi r_t \delta l/R_F^2$ due to the change in the path length traveled by the wave due to the new location of the patch; $r_t$ is the original distance of the coherent patch from the point in the screen where the observer-source line of sight intersects. For $\delta l\sim \ell_\pi$ and $r_t \sim R_{scat}/2$ (which is a typical value for a patch visible by the observer), the phase shift is $\sim \pi$ when $\ell_\pi > \ell_{min}$ (eq. \ref{rscat}). Thus, roughly half of the patches in the part of the screen visible to the observer introduce an additional phase-shift of order $\pi$ due to the transverse displacement of the screen by $\sim2\ell_\pi$. Therefore the scintillation pattern at the observer plane shifts by roughly one fringe width. If the relative transverse velocity between the scattering screen and the observer is $v_{os}$, then this second timescale is
	\begin{equation}
	\delta t_{ts} \sim { \max\left\{ \ell_\pi, \ell_{\rm min}\right\} \over 
		v_{\rm os}}. 
	\label{ext2}
	\end{equation}
	The fluctuation timescale for FRB light-curves due to propagation through a turbulent medium, $\delta t_{\rm var}$, is
	\begin{equation}
	\delta t_{\rm var}=\min(\delta t_{\rm ed}, \delta t_{\rm ts}) \sim 
	{ \max\left\{ \ell_\pi, \ell_{\rm min}\right\} \over \sqrt{v^2_{\rm ed} +
			v^2_{\rm os}} }.
	\end{equation}
	Light-curve variability due to scintillation requires $\delta t_{\rm var}<\max(\delta t,t_{\rm FRB})$. As we will show below this is only expected to happen 
	if the scattering screen is very close to the source.
	
	The phase difference between waves arriving
	at the observer from two points on the scattering screen separated by a 
	distance $\sim R_{\rm F}$ is $\pi$, 
	The phase difference between waves from the center of the screen and 
	radius $R_{\rm scat}$ is 
	$\sim\pi (R_{\rm scat}/R_{\rm F})^2$.
	Therefore, the flux at two frequencies separated by $\delta\nu$
	are uncorrelated when 
	\begin{equation}
	(\delta\nu) {d \over d\nu} \left[{\pi R^2_{scat}\over R^2_F}
	\right] \gae \pi \quad\implies\quad \delta\nu \gae 
	{\nu\, R^2_F\over R^2_{scat}}\sim {1\over \delta t},
	\label{delnu}
	\end{equation}
	where $\delta t$ is given by equation (\ref{del_t}). The timescale for the variation of spectrum is $\delta t_{\rm var}$. Weak scintillation occurs when 
	$R_{\rm scat}\lesssim R_{\rm F}$ and strong scintillation when 
	$R_{\rm scat}\gtrsim R_{\rm F}$. 
	
	We apply these results to wave propagation through turbulent inter-galactic 
	medium (IGM), FRB host galaxy and Milky Way ISM.
	We don't know whether the density fluctuations in the IGM and the host
	galaxies of FRBs follow the Kolmogoroff scaling. However, we know 
	that the spectrum in the Galaxy is Kolmogoroff spanning $10$-orders of 
	magnitude in length scale. Therefore, we will take the density 
	fluctuation index $\alpha=1/3$ (the Kolmogoroff value) for all numerical 
	estimates in the remainder of this section. The results can be easily 
	recalculated for a different index should observations provide that information.
	
	\subsection{Scattering in inter-galactic medium}
	\smallskip
	
	The electron density in the local IGM is $n_e\sim 10^{-7}$ cm$^{-3}$. The 
	size of the largest eddy in the IGM, $\ell_{\rm max}$, is highly uncertain 
	by several orders of magnitude. It might be as large as $10^{24}$cm, 
	which is the scale for energy deposition into the IGM by AGN jets and 
	outflows from galaxy clusters or as small as a few 10s pc.
	If the Mach number of IGM turbulence on the 
	largest eddy scale were to be order unity, then $\ell_{\rm max}$ cannot be
	much smaller than $10^{24}$cm. Otherwise, the heating of the IGM due to 
	dissipation of kinetic energy of turbulence would exceed the bremsstrahlung
	cooling rate, and the IGM temperature would rise on a timescale smaller than
	the Hubble time. This is contradicted by the data, which suggests that the
	IGM is heated by UV photons and its mean temperature of $\sim 10^4$K is 
	not increasing rapidly as the universe ages. The outer scale of turbulence
	can be smaller for lower Mach number turbulence as the constraint on
	$\ell_{\rm max}$ from turbulent heating of IGM scales as $M_t^3$.
	
	The thickness of the IGM {\it scattering screen} is of order the
	distance between the source and us, i.e. $L \sim d_{\rm so}\sim 10^9$pc.
	Therefore, the size of the smallest eddy for strong scattering,
	$\Delta\phi\sim 1$, at 1 GHz is estimated from equation (\ref{lpi}) 
	to be $\ell_\pi\sim 10^{15}$cm if we take $\ell_{\rm max}\sim 10^{24}$cm 
	and $\ell_\pi\sim 10^{14}$ cm for $\ell_{\rm max}\sim 10^{21}$cm.
	
	The Fresnel scale for IGM scatterings is $R_{\rm F}\sim 3\times 10^{14}$cm at 1 GHz. This is  marginally smaller than $\ell_\pi$ even in the extreme case of $\ell_{\rm max}\sim 10^{24}$ cm. Thus, the IGM scattering lies between the weak and strong scintillation regimes.
	The scattering would  be in the weak regime if the smallest eddies don't get down to the scale of $\ell_\pi$. We estimate the smallest scale for the turbulent cascade in the IGM.
	
	The smallest eddy size is the larger of the viscous dissipation length scale (eq. \ref{lmin1}) and the Larmor radius (eq. \ref{RB}) as long as the mean-free path is much larger than the Larmor radius.
	The measurement of IGM magnetic field is highly uncertain. Faraday rotation measurements of radio sources place an upper limit of $10^{-9}$G on IGM field with correlation length $>1$ Mpc. TeV photons from Blazars and GRBs interacting with the cosmic infrared background produce electron-positron pairs, and these pairs inverse-Compton scatter CMB photons to produce a secondary beam of GeV photons. The observed duration and angular width of the GeV pulse depends on the IGM magnetic field. Observations of TeV photons from Blazars and the follow up non-detection (upper limits) of the subsequent GeV shower provide a lower limit on the IGM field of $\sim 10^{-15}$G (with a large uncertainty) if the coherence length of the field, $l_{\rm mag}$, is larger than $1$\,Mpc, e.g. \citep{NV2010,Tavecchio2010}; the limit on the field strength scales as $l_{\rm mag}^{-1/2}$ for $l_{\rm mag} < 1$ Mpc.
	The expected field strength is $\sim 10^{-12}$G -- again with a large uncertainty -- if the IGM field were the frozen-in field in galactic outflows and AGN jets\footnote{We took the magnetic field strength to be $10^{-2}\mu$G in the outflow on a scale of 1 kpc. The transverse component of the frozen-in field falls off as $r^{-1}$ as the flow expands and at a distance of 10 Mpc the field is of order $10^{-12}$G.}.
	Thus, the proton Larmor radius, for IGM at temperature $10^4$K, is between $\sim 10^{14}$ \& $10^{17}$cm. The viscous length scale (eq.
	\ref{lmin1}), on the other hand, is $\sim10^{13}$cm ($10^{17}$cm) if the IGM magnetic field is 10$^{-12}$G (10$^{-15}$G). Thus, the size of the
	smallest eddy is expected to be between $\sim 10^{14}$ \& $10^{17}$cm depending on the IGM field. The smallest eddy is smaller than $\ell_\pi$ at the low end of this estimate, but otherwise the scattering angle for radio waves in the IGM is $<\lambda/\ell_\pi$.
	
	This suggests that the deflection angle for 1 GHz waves in the IGM is no larger than $\sim 3 \times 10^{-14}$rad (eq. \ref{deltheta}), and the corresponding temporal broadening of a pulse is $\delta t\lae 10^{-11}$s (eq. \ref{del_t}).  Moreover, the coherence bandwidth $\delta\nu/\nu\sim 1$. The light-curve variability due to IGM scatterings is on a time scale (for
	$\alpha=1/3$)
	\begin{eqnarray}
	\delta t_{\rm var}\!\approx \!\left\{ \begin{array}{ll}\hskip -6pt 10^8 {\rm s}\,
	{\ell_{\pi,15}\over v_{\rm max,7}} & \!\ell_{\pi}>\ell_{\rm min} ,\\ \\
	\hskip -6pt  10^{10}{\rm s}\, \!{\ell_{\rm min,17}\over v_{\rm max,7}} & \ell_{\pi}<\ell_{\rm min}.
	\end{array} \right.
	\end{eqnarray}
	where $v_{\rm max}\equiv \max(v_{\rm os},v_{\rm ed})$. The estimated timescale is much too long to be important for the ms duration FRBs.
	
	The line of sight to cosmological FRBs at a distance of a Gpc or more passes
	through several Lyman-alpha clouds. The electron density in these clouds
	is larger than the mean IGM density by a factor $\sim 10$, and thus
	radio waves are deflected when passing through these clouds by an angle
	that is a factor $\sim n_e^{1.2} L^{3/5}/\ell_{\rm max}^{2/5}\sim 10$ larger 
	than IGM scatterings; this is assuming that the smallest scale for 
	fluctuations in these clouds is $\lae \ell_\pi\sim 3\times 10^{13}$cm, and
	$\ell_{\rm max}\sim 10^2$pc, $L\sim 10^5$pc. GHz radio pulses are broadened while
	passing though these clouds by about 1 ns.
	
	The probability that our line of sight to a FRB at a distance of a few Gpc 
	passes through the outer halo of a galaxy or an intra-cluster medium is high.
	The electron density and the width of the medium are of order 
	10$^{-4}$cm$^{-3}$ and 1 Mpc respectively in this case. Taking the 
	largest scale for turbulence in this medium
	to be $\sim 0.1$ Mpc, we find the deflection angle and pulse broadening
	to be $\sim 10^{-12}$rad and $10^{-2}\mu$s.
	
	The bottom line is that the FRB pulses are broadened the least ($\lae 10^{-2}$ns) while passing though the turbulent IGM plasma and the most (a few ns) by the intra-cluster medium (the probability for encountering which at $1$\,Gpc is a few percent). The coherence bandwidth due to scatterings for all the cases considered in this sub-section is $\delta\nu/\nu \sim 1$.
	
	\subsection{Scattering in the Milky Way and FRB host galaxy ISM}
	\smallskip
	
	An excellent approximation for the Fresnel scale for a scattering screen in the Milky Way galaxy or the FRB host galaxy is (eq. \ref{Rf})
	\begin{equation}
	R_{\rm F} \approx (\lambda d)^{1/2} = (3\times 10^{11}\,{\rm cm})\, 
	\nu_9^{1/2} d_{\rm kpc}^{1/2}, 
	\end{equation}
	where $\nu_9$ is the wave frequency in GHz, $d$ is the distance between the 
	scattering screen and the FRB source or the screen and the observer whichever 
	is smaller, $d_{\rm kpc}$ is $d$ in units of 1 kpc. The size of the smallest 
	eddies for strong scattering ($\ell_\pi$), the diffraction scale, is obtained 
	from equation (\ref{lpi}) for Kolmogoroff density fluctuations ($\alpha=1/3$)
	\begin{multline}
	\label{eq:lpiISM}
	\quad\quad\ell_\pi \sim (2{\rm x}10^{13} {\rm cm})\, n_e^{-{6\over5}} L^{-{1\over 5}}
	\nu_9^{6\over 5} \left( {\ell_{max}\over L}\right)^{2\over 5} \\
	\sim (4{\rm x}10^{9} {\rm cm})\, {\rm DM}_{\rm s}^{-{6\over5}} 
	\nu_9^{6\over 5} \left( {\ell_{max}\over L}\right)^{2\over 5} 
	{ L\over 1 {\rm pc}}, \;\,\quad\quad
	\end{multline}
	where ${\rm DM}_{\rm s}=n_e L$ is the contribution to the dispersion measure (DM) from electrons in the scattering screen (measured in $\mbox{pc cm}^{-3}$). The diffraction scale is larger than the smallest eddy size $\sim 10^8$cm for 1 $\mu$G magnetic field and 10$^4$K temperature of the medium (eqs. \ref{lmin2}). Thus, the scattering is in the strong regime and we can consider the special case of scintillation where $\ell_{\rm min} < \ell_\pi$.
	
\begin{figure}
		\centering
		\includegraphics[width = .45\textwidth]{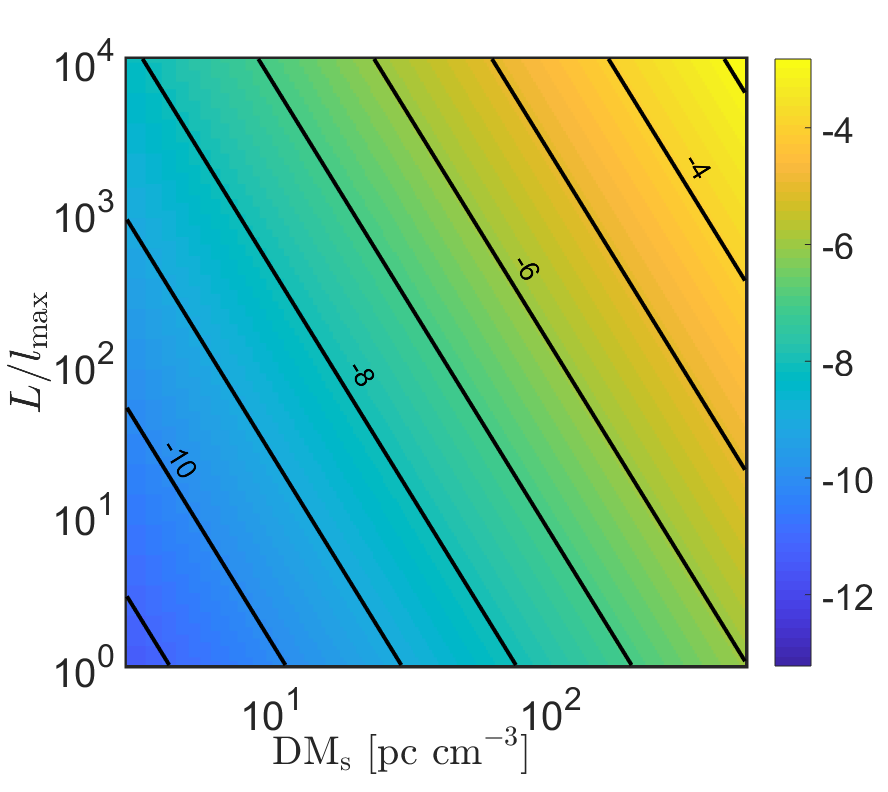}
		\caption{  
			Pulse broadening ($\delta t$) at $1$\,GHz due to a scattering screen at a distance $d=1$\,kpc from the FRB source is shown in this figure as a function of the electron column density associated with the screen ($\mbox{DM}_{\rm s}$) and $L/\ell_{\rm max}$ which is the ratio of the  width of the screen along the line of sight to the observer and the size of the largest turbulent eddies. The labels on the contours are ${\log}_{10}( \delta t)$ in seconds. For a screen at a redshift $z$ these numbers should be divided by $(1+z)^{3.4}$, and for a scattering screen at a distance $d$ (expressed in parsec) from the FRB, $\delta t$ shown in the graph changes by a factor $10^3/d$.  Results for a scattering screen in our galaxy are the same as in this figure when we take the distance $d$ to be between the screen and us. 
		} 
		\label{deltat}
	\end{figure}

\begin{figure*}
		\centering
		\includegraphics[width = .45\textwidth]{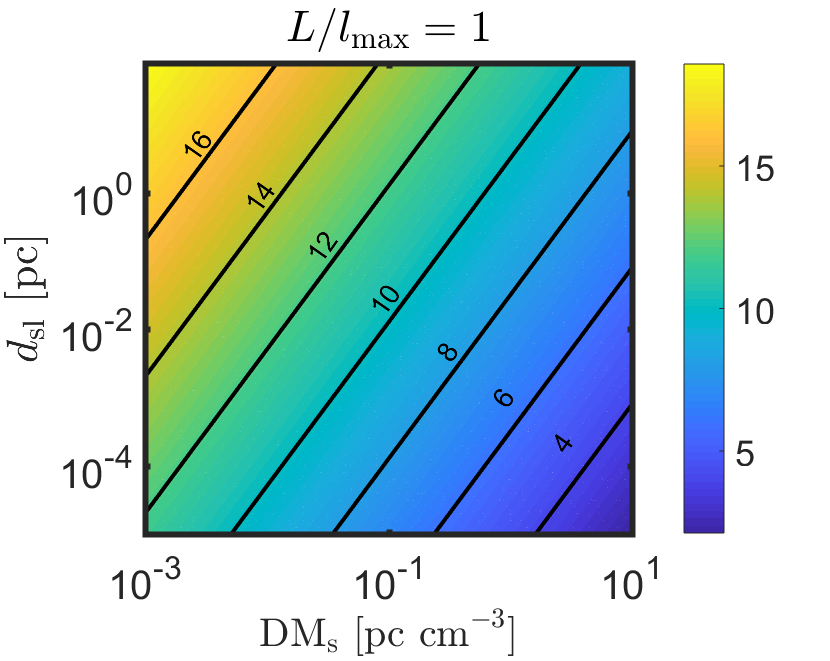}
		\includegraphics[width = .45\textwidth]{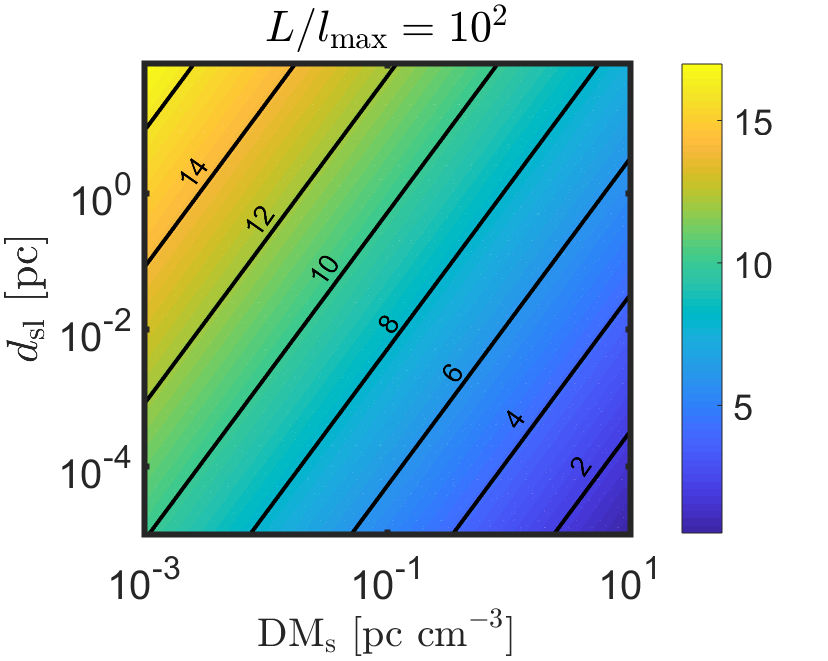}\\
		\includegraphics[width = .45\textwidth]{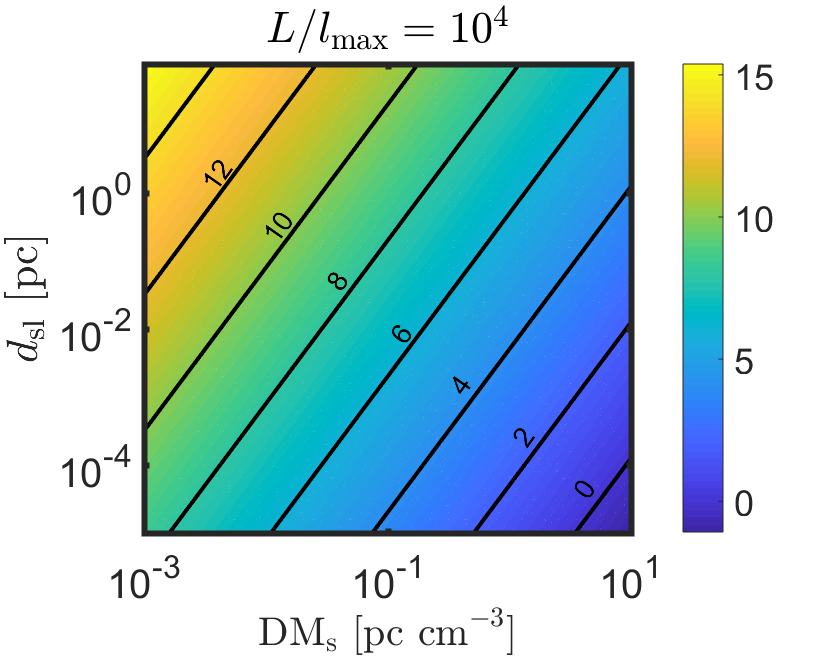}
		\includegraphics[width = .45\textwidth]{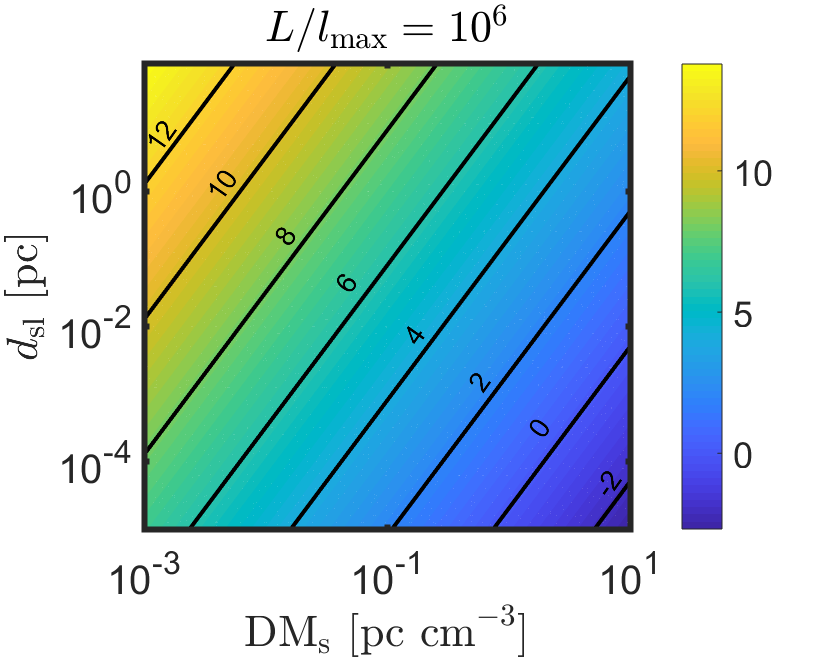}
		\caption{  
			{Coherence bandwidth (Hz) at $1$\,GHz due to a scattering screen close to the 
				source. Results are shown as a function of the screen's contribution to the 
				dispersion measure ($\mbox{DM}_{\rm s}$), the screen's distance from the 
				source ($d_{\rm sl}$) and the ratio of the screen thickness and the largest 
				eddy size ($L/l_{\rm max}$). The black lines are curves of constant coherence 
				bandwidth, and the number associated with each line is Log$_{10}$ of the
				coherence bandwidth in Hz.}
		} 
		\label{coband}
	\end{figure*}

	The ISM of Milky Way consists of multiple phases and the electron density in these phases varies by several orders of magnitude. The electron density of the ISM of FRB host galaxy and in the near vicinity of the object is largely unknown. Given this uncertainty, it is better to parametrize the pulse smearing by scintillation in terms of parameters as closely related to observables as we possibly can. One such parameter is the dispersion measure (DM). Although we measure only the total DM for FRBs,
	and not contributions from the ionized nebula surrounding the source, host galaxy ISM, and various other components separately, we can at least
	place an upper bound on the contributions from these components. We can rewrite the equation for pulse broadening (eq. \ref{del_t}) in terms of the dispersion measure in the scattering screen, $\mbox{DM}_{\rm s}$, as follows:
	\begin{equation}
	\delta t \sim \mbox{DM}_{\rm s}^{4\over 2\alpha+1}\, 
	\left({q^2\lambda\, \mbox{pc}\over m_e\,c^2}\right)^{4\over 2\alpha+1} 
	{\lambda^2\over 2\pi^2c d}\left[ {d\over L}\right]^2 \left[ {L\over \ell_{\rm max}}
	\right]^{4\alpha\over 2\alpha+1},
	\label{dt5}
	\end{equation}
	where $\mbox{pc}=3.1\times 10^{18}$cm is one parsec in cm, $L$ is the width of the scattering screen which for most situations is expected to be of order $d$, and
	$\mbox{DM}_{\rm s}$ is measured in the units of pc cm$^{-3}$.
	The factor $L/\ell_{\rm max}$ is likely to have a large uncertainty as we don't
	know the scale for energy injection in the turbulent screen. The above
	expression for $\delta t$ should be divided by $(1+z)^{{2\alpha+5\over 
			2\alpha+1}}$ when the scattering is in the FRB host galaxy at
	redshift $z$. 
	For the special case of Kolmogoroff turbulence, $\alpha=1/3$, we can 
	write (\ref{dt5}) in the following more convenient form for observational use
	\begin{equation}
	\delta t \sim (8\times 10^{-13}{\rm s})\, \nu_9^{-4.4} 
	(1+z)^{-3.4} {\mbox{DM}_{\rm s}^{2.4}\over d_{\rm kpc}} \left({d\over L}\right)^2 \,
	\left[ {L\over \ell_{\rm max}}\right]^{0.8}.
	\label{dt7}
	\end{equation}
	We see that $\delta t$ has a strong dependence on $\mbox{DM}_{\rm s}$, an almost linear dependence on $L/\ell_{\rm max}$ and it scales inversely with $d$
	(distance of the screen from the source or the observer, whichever is smaller).
	For a scattering screen in the FRB host galaxy at redshift 1, at a distance of 1 kpc from the source, which has $L/\ell_{\rm max}\sim 10^3$ and $\mbox{DM}_{\rm s}\sim 10^2$, we find that $\delta t \sim 1 \mu$s  at $1$\,GHz. A plasma screen at a distance of 0.1 pc from the source, that has $\mbox{DM}_{\rm s}\sim 10\mbox{ pc cm}^{-3}$ and $L/\ell_{\rm max}\sim 10^2$, gives $\delta t\sim 7\mu$s.
	We show in figure \ref{deltat} a contour plot of $\delta t$ as a function of $\mbox{DM}_{\rm s}$ and $L/\ell_{\rm max}$ that provides a quick estimate of the parameters that can account for the measured temporal broadening of FRB light-curves. The coherence bandwidth of the spectrum is $\sim 1/\delta t$. The coherence bandwidth ($\delta\nu$), given by eq. \ref{delnu}, for different screen 
	parameters is shown in figure \ref{coband}; a coherence bandwidth $\gtrsim$\,MHz at $\nu = 1$ GHz generally requires a scattering screen that is relatively close to the source (or the observer) so that the ${\rm DM}_{\rm s}\ll$ DM.
	The variability time due to scintillation is
	\begin{eqnarray}
	\delta t_{\rm var}\approx 100\, \ell_{\pi,9}v_{\rm max,7}^{-1}\mbox{ s}.
	\end{eqnarray}
	This is much larger than the FRB burst duration unless the screen is extremely close to the source (or observer), causing $l_{\pi}$ to decrease (see eq. \ref{eq:lpiISM}). For example, in order for the variability to be of the order of $\sim 1$\,ms (while keeping $\mbox{DM}_{\rm s}$ fixed at $\mbox{DM}_{\rm s}\sim 1\mbox{pc cm}^{-3}$) one requires $d\sim L\sim 10^{13}$cm. Note however that the linear theory of scattering and pulse broadening breaks down within a distance of $10^{12} L_{frb,40}^{1/2}$ cm of the FRB source where the wave nonlinearity parameter is $\gae 1$; $L_{frb,40}$ is FRB luminosity at 1 GHz in units of 10$^{40}$ erg s$^{-1}$. Some of the nonlinear effects can be handled using the work of \cite{LuPhinney2020}.
	
	\bigskip

	\section{Summary and discussion}
	
	The variability of FRB light-curves and any break in the spectrum contain information regarding the radiation mechanism, which has not been made use of thus far. This work describes what we can hope to learn from light-curve variability and spectral features about FRB physics.
	
	We have considered the scenario where the FRB emission is produced at some distance $R$ from the magnetar by a relativistic outflow that is moving toward the observer with Lorentz factor $\gamma\gg 1$. The duration of the outflow in the magnetar rest frame is $t_{jet}$.
	As long as $t_{jet}$ is much larger than $t_0 = R/(2c \gamma^2)$, the observed duration of the FRB, $t_{\rm FRB}$ is dictated by $t_{\rm FRB}\sim t_{jet}$ and the FRB light-curve reflects the temporal structure of the outflow. In this case, the light-curve can turn on and off quickly compared with the duration of the pulse, and the timescale for temporal fluctuations of the light-curve ($\sim t_0$) can be very rapid\footnote{If the intrinsic variability time is found to be a few $\mu$s or smaller then that would be a good indication that FRB radiation is produced in the close vicinity of a compact object, well inside the magnetosphere, as suggested by the model of \citep{Kumar+17} which is further developed in \cite{KumarBosnjak2020}.}. Similarly, the spectrum can have sharp structures when $R\ll 10^{10}$cm.
	
	However, when the radio emission is produced outside the magnetar magnetosphere such that $t_{\rm FRB}\sim t_0$, then the rise time of the light-curve, its decline, temporal fluctuations, and the spectral features are all highly constrained 
	by the geometry of the shock front and special 
	relativity. Much of the paper, and the discussion here, addresses what FRB data can tell us about the 
	viability of this class of models. A special case of this general scenario we have considered is maser emission in the shock driven by the relativistic outflow into the circum-stellar medium of the magnetar.
	
	For FRB emission produced outside the magnetosphere, the natural timescale for the rise of the light-curve is $t_0$ (defined above), which is also of order the duration of a pulse in the FRB light-curve.
	Thus, we expect the ratio of the rise time and the pulse duration to be of order unity; the expected ratio of the light-curve variability time and the pulse duration is also of order unity in this case. One way to get these ratios to be much less than one is by concentrating the maser emission process to a small patch of the shocked plasma of comoving size 
	$\zeta R/\gamma$ with $\zeta\ll 1$, i.e. the emission is produced in an area of size much smaller than the shock front surface visible to the observer. However, in this case the efficiency of 
	radio production is reduced by a factor $\sim \zeta^2-\zeta$ (\S2.3.1) above and beyond the efficiency of the maser process. Alternative ways to get a rapid rise time compared to the pulse duration involve producing the emission in a very narrow range of radii or when the spectrum is very narrow in the comoving frame, and it rapidly sweeps through the observed band. These scenarios too, lead to a significantly reduced efficiency. Perhaps the most promising way to obtain rapid fluctuations of the light-curve is by arranging the emission to be highly anisotropic in the comoving frame of the source (see \S\ref{sec:anisotropic}); the FRB pulse shape in this case should be double peaked, or horn shaped, as shown in Fig. \ref{fig:var}.
	
	The fastest rate for the decline of light-curves is set by the emission from outside the relativistic beaming angle, or parts of the outflow at angles lager than $\gamma^{-1}$ wrt to the observer line of sight, that arrives at the observer. Thus, the fastest possible decline of the light-curve at a fixed frequency is: $f_{\nu} \propto t^{-2-\tilde{\beta}}$ \citep{KP2000}; where $\tilde{\beta}$ is the spectral index defined as $f_\nu\propto\nu^{-\tilde{\beta}}$, and $f_\nu$ is the specific flux. In \S \ref{sec:temproscetral} we extended this result to spectra that may be intrinsically narrow (in frequency and/or time), and for observations done at different frequencies. One of the key results we found is that if the observed specific flux at $\nu_1,t_1$ is $f_1$ then the flux at a lower frequency $\nu_2$, and at a later time, $t_2=t_1 \nu_1/\nu_2$, should not be smaller than $f_1(\nu_2/\nu_1)^2$ for the class of models where $t_{\rm FRB} \sim t_0$.
	In other words, the fall-off of spectrum at low frequencies that is faster than $\nu^2$ is inconsistent with the expectation of radiation being produced at $R\gtrsim 10^{10}$cm. 
	These results regarding the temporal fluctuations of the light-curves and spectra are purely geometric in origin and follow from special relativity, 
	and are largely independent of the details of the radiation mechanism\footnote{The 
		main assumption in this calculation is that the emitting region is optically thin to the observed radiation. The most relevant optical depth to consider is due to induced Compton scattering. By construction, radiation that reaches the observer must be at least moderately optically thin to this process for the majority of photons, which are produced within $\gamma^{-1}$ from the line of sight. For photons produced at higher latitudes and traveling toward the observer, the induced Compton optical depth is expected to only be lower, due to a reduction in the occupation number of photons produced outside of the relativistic beaming cone. Therefore, Induced Compton cannot easily suppress high latitude photons.}
	as long as the process involves a relativistic outflow of angular size larger than $\gamma^{-1}$ outside the magnetosphere\footnote{A faster decline of the flux, either in time or frequency, can arise provided that the angular size of the outflow is $\lae\gamma^{-1}$.}. One notable caveat is that radiation produced at $\theta>1/\gamma$ re-encounters the emitting shell before reaching the observer. The frequency of the photon at this second encounter with the shell is smaller  in the local comoving frame. The high-latitude radiation therefore may be suppressed at this second encounter if its comoving frame frequency is smaller than the local plasma frequency. However, it is important to note that this can only suppress the high-latitude signature if the intrinsic spectrum is both narrow and centered close to the local plasma frequency. Even in this scenario, the radiation is still visible from photons emitted at least up to an angle of $1/\gamma_0$ and so the observed spectral and temporal widths of the signal will be at least of order unity.
	
	Of course, scintillation effects should be removed from the data before checking for the steepness of the spectrum according to the argument above.
	Radio scintillation in the FRB host galaxy, IGM, and our galaxy can smooth out the intrinsic variability of FRB light-curves and introduce features in their spectra. In \S \ref{sec:Extvar} we estimated the inner scale of turbulence due to viscous damping, and provide a formula for scintillation time as a function of electron column density of the scattering screen, its distance from the 
	source (or observer) and the outer scale of the turbulence, to help determine intrinsic FRB properties.
	
\begin{figure}
		\centering
		\includegraphics[width = .45\textwidth]{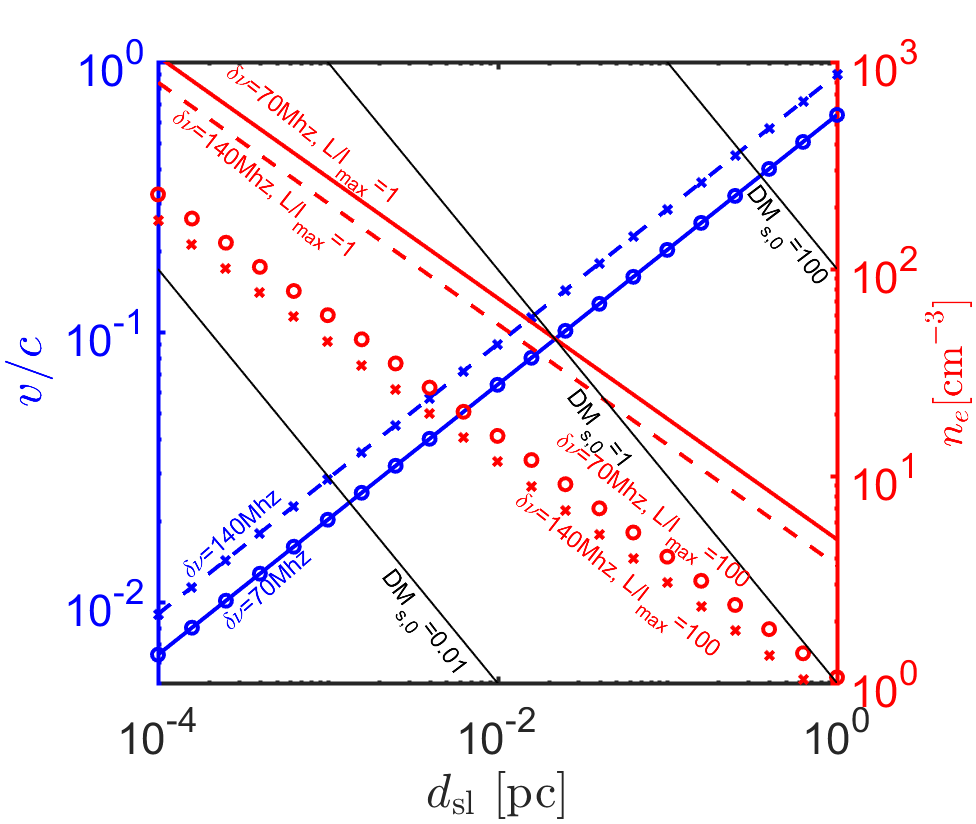}
		\caption{ Shown here is the transverse velocity of the scattering screen, or the speed of turbulence on the diffraction scale, as a function of the distance of the scattering screen from the FRB source in order that the scintillation time is $\delta t_{\rm var}=30$\,ms and the coherence bandwidth for scintillation 
			($\delta\nu$ -- eq. \ref{delnu}) is 70 MHz or 140 MHz; the velocity is independent of the parameter $L/\ell_{max}$. We see that a modest speed, $v/c\lae 10^{-2}$, can give $\delta t_{var} = 30$ ms \& $\delta\nu\sim 10^2$MHz when the screen is at a distance $\sim10^{-4}$pc from the source. Also shown is the electron density in the scattering screen for two different values of $L/\ell_{max}$ (the right-hand side of the y-axis is the density scale), and the contribution of the screen to the DM of the FRB source (where $\mbox{DM}_{\rm s,0}\equiv \mbox{DM}_{\rm s}/\mbox{pc cm}^{-3}$).}
		\label{fig:coherdelnu}
	\end{figure}

	A specific example of an application of the results presented in this work is the case of the Galactic FRB 200428 (\S\ref{sec:intvar}). The spectrum for the second pulse of this burst was cutoff abruptly below $\sim 550$MHz.
	The first radio pulse of FRB 200428, which preceded the second pulse by $30$\,ms, was detected only between $400$\,MHz and $550$\,MHz by CHIME. Scintillation can cause the spectrum to change in $30$\,ms with a coherence bandwidth of $\sim 10^2$MHz at $\nu\sim 1$ GHz provided that the scattering screen is within a few parsecs of the source (see Fig. \ref{fig:coherdelnu}); the required transverse velocity of the screen wrt to the source-observer line of sight is $\lae 10^{-2}$c when the screen is at a distance $\lae 10^{-3}$pc from the source\footnote{Plasma with fluctuating electron density at a distance $\sim 10^{-3}$pc from the magnetar is certainly plausible. This view is supported by observations that FRB producing neutron stars, such as SGR 1935+2154 in our galaxy, have many outbursts on time scales of minutes to days and probably have unsteady wind as well. The outflow/wind at a distance of $10^{-3}$ pc from the NS becomes cold due to adiabatic expansion and density fluctuations are not likely to be wiped out.} (Fig. \ref{fig:coherdelnu}) and the size of the source is $\ll 10^9$cm such as when the radiation is produced inside the NS magnetosphere \citep{LKZ2020}. However, the fact that the first pulse was not detected between 1281 \& 1468 MHz \citep{STARE2020}, and most likely the spectrum was cutoff above 550 MHz, suggests an intrinsically narrow spectrum for this pulse. The low-frequency cutoff of the spectrum of the second pulse is also likely to be intrinsic to the source and not due to scintillation; it would require a high degree of fine tuning of scintillation parameters to produce the sharp high and low frequency cutoffs for the first and second pulses and the flux to be unobservably small over several hundred MHz at least. The presence of an intrinsic cutoff of the spectrum, for the second pulse in particular, would rule out the FRB model in which radio emission is produced outside the NS light-cylinder, independent of other difficulties specific to the shock model pointed out by \cite{LKZ2020}.
	
	Another FRB observation which is marginally in conflict with the basic expectations of radiation being produced at $R\gtrsim 10^{10}$cm comes from FRB 121102. Three bursts from this FRB were detected by VLA at 2.5-3.5 GHz but not by Arecibo at 1.15-1.73 GHz, despite Arecibo being more sensitive than VLA by a factor $\sim 5$ \citep{Law+17}. Given that the Arecibo band is a factor $\sim 2$ lower than the VLA, the expected flux from a relativistic outflow in the Arecibo band should have been at most $\sim 4$ times less than the VLA flux if the angular size of the outflow is larger than $\gamma^{-1}$.

	A number of current FRB observatories such as CHIME, ASKAP, DSA, have micro-second or better time resolution and it should be possible to determine fluctuations in radio light-curves down to those timescales for bright bursts. The power density spectrum of FRB light-curves in the frequency range of a few KHz to tens of MHz would provide important information regarding the FRB mechanism. Similarly, sharp features in the spectra, which are not due to scintillation, are good diagnostic tools of the FRB radiation mechanism.
	
	{\bf Data availability} The data produced in this study will be shared on reasonable request to the authors.
	
	\section{acknowledgments}
	PB thanks Wenbin Lu and Ben Margalit for helpful discussions. The research of PB was funded by the Gordon and Betty Moore Foundation through Grant GBMF5076. This work has been funded in part by an NSF grant AST-2009619.
	

\begin{thebibliography}{}
	\makeatletter
	\relax
	\def\mn@urlcharsother{\let\do\@makeother \do\$\do\&\do\#\do\^\do\_\do\%\do\~}
	\def\mn@doi{\begingroup\mn@urlcharsother \@ifnextchar [ {\mn@doi@}
		{\mn@doi@[]}}
	\def\mn@doi@[#1]#2{\def\@tempa{#1}\ifx\@tempa\@empty \href
		{http://dx.doi.org/#2} {doi:#2}\else \href {http://dx.doi.org/#2} {#1}\fi
		\endgroup}
	\def\mn@eprint#1#2{\mn@eprint@#1:#2::\@nil}
	\def\mn@eprint@arXiv#1{\href {http://arxiv.org/abs/#1} {{\tt arXiv:#1}}}
	\def\mn@eprint@dblp#1{\href {http://dblp.uni-trier.de/rec/bibtex/#1.xml}
		{dblp:#1}}
	\def\mn@eprint@#1:#2:#3:#4\@nil{\def\@tempa {#1}\def\@tempb {#2}\def\@tempc
		{#3}\ifx \@tempc \@empty \let \@tempc \@tempb \let \@tempb \@tempa \fi \ifx
		\@tempb \@empty \def\@tempb {arXiv}\fi \@ifundefined
		{mn@eprint@\@tempb}{\@tempb:\@tempc}{\expandafter \expandafter \csname
			mn@eprint@\@tempb\endcsname \expandafter{\@tempc}}}
	
	\bibitem[\protect\citeauthoryear{{Babul} \& {Sironi}}{{Babul} \&
		{Sironi}}{2020}]{Babul2020}
	{Babul} A.-N.,  {Sironi} L.,  2020, arXiv e-prints, \href
	{https://ui.adsabs.harvard.edu/abs/2020arXiv200603081B} {p. arXiv:2006.03081}
	
	\bibitem[\protect\citeauthoryear{{Bannister} et~al.,}{{Bannister}
		et~al.}{2017}]{Bannister2017}
	{Bannister} K.~W.,  et~al., 2017, \mn@doi [\apjl] {10.3847/2041-8213/aa71ff},
	\href {https://ui.adsabs.harvard.edu/abs/2017ApJ...841L..12B} {841, L12}
	
	\bibitem[\protect\citeauthoryear{{Bannister} et~al.,}{{Bannister}
		et~al.}{2019}]{Bannister+19}
	{Bannister} K.~W.,  et~al., 2019, \mn@doi [Science] {10.1126/science.aaw5903},
	\href {https://ui.adsabs.harvard.edu/abs/2019Sci...365..565B} {365, 565}
	
	\bibitem[\protect\citeauthoryear{{Beloborodov}}{{Beloborodov}}{2017}]{Beloborodov17}
	{Beloborodov} A.~M.,  2017, \mn@doi [\apjl] {10.3847/2041-8213/aa78f3}, \href
	{https://ui.adsabs.harvard.edu/abs/2017ApJ...843L..26B} {843, L26}
	
	\bibitem[\protect\citeauthoryear{{Beniamini} \& {Granot}}{{Beniamini} \&
		{Granot}}{2016}]{BG2016}
	{Beniamini} P.,  {Granot} J.,  2016, \mn@doi [\mnras] {10.1093/mnras/stw895},
	\href {https://ui.adsabs.harvard.edu/abs/2016MNRAS.459.3635B} {459, 3635}
	
	\bibitem[\protect\citeauthoryear{{Beniamini}, {Wadiasingh}  \&
		{Metzger}}{{Beniamini} et~al.}{2020}]{Beniamini+20}
	{Beniamini} P.,  {Wadiasingh} Z.,   {Metzger} B.~D.,  2020, \mn@doi [\mnras]
	{10.1093/mnras/staa1783}, \href
	{https://ui.adsabs.harvard.edu/abs/2020MNRAS.496.3390B} {496, 3390}
	
	\bibitem[\protect\citeauthoryear{{Bochenek}, {Ravi}, {Belov}, {Hallinan},
		{Kocz}, {Kulkarni}  \& {McKenna}}{{Bochenek} et~al.}{2020}]{STARE2020}
	{Bochenek} C.~D.,  {Ravi} V.,  {Belov} K.~V.,  {Hallinan} G.,  {Kocz} J.,
	{Kulkarni} S.~R.,   {McKenna} D.~L.,  2020, arXiv e-prints, \href
	{https://ui.adsabs.harvard.edu/abs/2020arXiv200510828B} {p. arXiv:2005.10828}
	
	\bibitem[\protect\citeauthoryear{{CHIME/FRB Collaboration} et~al.,}{{CHIME/FRB
			Collaboration} et~al.}{2019a}]{CHIME2019}
	{CHIME/FRB Collaboration} et~al., 2019a, \mn@doi [\nat]
	{10.1038/s41586-018-0867-7}, \href
	{https://ui.adsabs.harvard.edu/abs/2019Natur.566..230C} {566, 230}
	
	\bibitem[\protect\citeauthoryear{{CHIME/FRB Collaboration} et~al.,}{{CHIME/FRB
			Collaboration} et~al.}{2019b}]{CHIME2019b}
	{CHIME/FRB Collaboration} et~al., 2019b, \mn@doi [\nat]
	{10.1038/s41586-018-0864-x}, \href
	{https://ui.adsabs.harvard.edu/abs/2019Natur.566..235C} {566, 235}
	
	\bibitem[\protect\citeauthoryear{{Chatterjee} et~al.,}{{Chatterjee}
		et~al.}{2017}]{Chatterjee+17}
	{Chatterjee} S.,  et~al., 2017, \mn@doi [\nat] {10.1038/nature20797}, \href
	{https://ui.adsabs.harvard.edu/abs/2017Natur.541...58C} {541, 58}
	
	\bibitem[\protect\citeauthoryear{{Cho} et~al.,}{{Cho} et~al.}{2020}]{Cho2020}
	{Cho} H.,  et~al., 2020, \mn@doi [\apjl] {10.3847/2041-8213/ab7824}, \href
	{https://ui.adsabs.harvard.edu/abs/2020ApJ...891L..38C} {891, L38}
	
	\bibitem[\protect\citeauthoryear{{Cordes}, {Wasserman}, {Hessels}, {Lazio},
		{Chatterjee}  \& {Wharton}}{{Cordes} et~al.}{2017}]{Cordes2017}
	{Cordes} J.~M.,  {Wasserman} I.,  {Hessels} J.~W.~T.,  {Lazio} T.~J.~W.,
	{Chatterjee} S.,   {Wharton} R.~S.,  2017, \mn@doi [\apj]
	{10.3847/1538-4357/aa74da}, \href
	{https://ui.adsabs.harvard.edu/abs/2017ApJ...842...35C} {842, 35}
	
	\bibitem[\protect\citeauthoryear{{Farah} et~al.,}{{Farah}
		et~al.}{2018}]{Farah2018}
	{Farah} W.,  et~al., 2018, \mn@doi [\mnras] {10.1093/mnras/sty1122}, \href
	{https://ui.adsabs.harvard.edu/abs/2018MNRAS.478.1209F} {478, 1209}
	
	\bibitem[\protect\citeauthoryear{{Gajjar} et~al.,}{{Gajjar}
		et~al.}{2018}]{Gajjar2018}
	{Gajjar} V.,  et~al., 2018, \mn@doi [\apj] {10.3847/1538-4357/aad005}, \href
	{https://ui.adsabs.harvard.edu/abs/2018ApJ...863....2G} {863, 2}
	
	\bibitem[\protect\citeauthoryear{{Ghisellini} \& {Locatelli}}{{Ghisellini} \&
		{Locatelli}}{2018}]{Ghisellini2018}
	{Ghisellini} G.,  {Locatelli} N.,  2018, \mn@doi [\aap]
	{10.1051/0004-6361/201731820}, \href
	{https://ui.adsabs.harvard.edu/abs/2018A&A...613A..61G} {613, A61}
	
	\bibitem[\protect\citeauthoryear{{Goldreich} \& {Sridhar}}{{Goldreich} \&
		{Sridhar}}{1995}]{Goldreich1995}
	{Goldreich} P.,  {Sridhar} S.,  1995, \mn@doi [\apj] {10.1086/175121}, \href
	{https://ui.adsabs.harvard.edu/abs/1995ApJ...438..763G} {438, 763}
	
	\bibitem[\protect\citeauthoryear{{Hessels} et~al.,}{{Hessels}
		et~al.}{2019}]{Hessels+19}
	{Hessels} J.~W.~T.,  et~al., 2019, \mn@doi [\apjl] {10.3847/2041-8213/ab13ae},
	\href {https://ui.adsabs.harvard.edu/abs/2019ApJ...876L..23H} {876, L23}
	
	\bibitem[\protect\citeauthoryear{{Katz}}{{Katz}}{2014}]{Katz2014}
	{Katz} J.~I.,  2014, \mn@doi [\prd] {10.1103/PhysRevD.89.103009}, \href
	{https://ui.adsabs.harvard.edu/abs/2014PhRvD..89j3009K} {89, 103009}
	
	\bibitem[\protect\citeauthoryear{{Katz}}{{Katz}}{2016}]{Katz16}
	{Katz} J.~I.,  2016, \mn@doi [\apj] {10.3847/0004-637X/826/2/226}, \href
	{https://ui.adsabs.harvard.edu/abs/2016ApJ...826..226K} {826, 226}
	
	\bibitem[\protect\citeauthoryear{{Katz}}{{Katz}}{2018}]{Katz2018}
	{Katz} J.~I.,  2018, \mn@doi [Progress in Particle and Nuclear Physics]
	{10.1016/j.ppnp.2018.07.001}, \href
	{https://ui.adsabs.harvard.edu/abs/2018PrPNP.103....1K} {103, 1}
	
	\bibitem[\protect\citeauthoryear{{Katz}}{{Katz}}{2020}]{Katz2020}
	{Katz} J.~I.,  2020, arXiv e-prints, \href
	{https://ui.adsabs.harvard.edu/abs/2020arXiv200603468K} {p. arXiv:2006.03468}
	
	\bibitem[\protect\citeauthoryear{{Kocz} et~al.,}{{Kocz}
		et~al.}{2019}]{Kocz2019}
	{Kocz} J.,  et~al., 2019, \mn@doi [\mnras] {10.1093/mnras/stz2219}, \href
	{https://ui.adsabs.harvard.edu/abs/2019MNRAS.489..919K} {489, 919}
	
	\bibitem[\protect\citeauthoryear{{Kumar} \& {Bo{\v{s}}njak}}{{Kumar} \&
		{Bo{\v{s}}njak}}{2020}]{KumarBosnjak2020}
	{Kumar} P.,  {Bo{\v{s}}njak} {\v{Z}}.,  2020, \mn@doi [\mnras]
	{10.1093/mnras/staa774}, \href
	{https://ui.adsabs.harvard.edu/abs/2020MNRAS.494.2385K} {494, 2385}
	
	\bibitem[\protect\citeauthoryear{{Kumar} \& {Panaitescu}}{{Kumar} \&
		{Panaitescu}}{2000}]{KP2000}
	{Kumar} P.,  {Panaitescu} A.,  2000, \mn@doi [\apjl] {10.1086/312905}, \href
	{https://ui.adsabs.harvard.edu/abs/2000ApJ...541L..51K} {541, L51}
	
	\bibitem[\protect\citeauthoryear{{Kumar}, {Lu}  \& {Bhattacharya}}{{Kumar}
		et~al.}{2017}]{Kumar+17}
	{Kumar} P.,  {Lu} W.,   {Bhattacharya} M.,  2017, \mn@doi [\mnras]
	{10.1093/mnras/stx665}, \href
	{https://ui.adsabs.harvard.edu/abs/2017MNRAS.468.2726K} {468, 2726}
	
	\bibitem[\protect\citeauthoryear{{Law} et~al.,}{{Law} et~al.}{2017}]{Law+17}
	{Law} C.~J.,  et~al., 2017, \mn@doi [\apj] {10.3847/1538-4357/aa9700}, \href
	{https://ui.adsabs.harvard.edu/abs/2017ApJ...850...76L} {850, 76}
	
	\bibitem[\protect\citeauthoryear{{Li} et~al.,}{{Li} et~al.}{2020}]{Li2020}
	{Li} C.~K.,  et~al., 2020, arXiv e-prints, \href
	{https://ui.adsabs.harvard.edu/abs/2020arXiv200511071L} {p. arXiv:2005.11071}
	
	\bibitem[\protect\citeauthoryear{{Lorimer}, {Bailes}, {McLaughlin}, {Narkevic}
		\& {Crawford}}{{Lorimer} et~al.}{2007}]{Lorimer+07}
	{Lorimer} D.~R.,  {Bailes} M.,  {McLaughlin} M.~A.,  {Narkevic} D.~J.,
	{Crawford} F.,  2007, \mn@doi [Science] {10.1126/science.1147532}, \href
	{https://ui.adsabs.harvard.edu/abs/2007Sci...318..777L} {318, 777}
	
	\bibitem[\protect\citeauthoryear{{Lu} \& {Kumar}}{{Lu} \&
		{Kumar}}{2018}]{LuKumar2018}
	{Lu} W.,  {Kumar} P.,  2018, \mn@doi [\apj] {10.3847/1538-4357/aad54a}, \href
	{https://ui.adsabs.harvard.edu/abs/2018ApJ...865..128L} {865, 128}
	
	\bibitem[\protect\citeauthoryear{{Lu} \& {Phinney}}{{Lu} \&
		{Phinney}}{2020}]{LuPhinney2020}
	{Lu} W.,  {Phinney} E.~S.,  2020, \mn@doi [\mnras] {10.1093/mnras/staa1679},
	\href {https://ui.adsabs.harvard.edu/abs/2020MNRAS.tmp.1797L} {}
	
	\bibitem[\protect\citeauthoryear{{Lu}, {Kumar}  \& {Zhang}}{{Lu}
		et~al.}{2020}]{LKZ2020}
	{Lu} W.,  {Kumar} P.,   {Zhang} B.,  2020, arXiv e-prints, \href
	{https://ui.adsabs.harvard.edu/abs/2020arXiv200506736L} {p. arXiv:2005.06736}
	
	\bibitem[\protect\citeauthoryear{{Lyubarsky}}{{Lyubarsky}}{2005}]{Lyubarsky2005}
	{Lyubarsky} Y.~E.,  2005, \mn@doi [\mnras] {10.1111/j.1365-2966.2005.08767.x},
	\href {https://ui.adsabs.harvard.edu/abs/2005MNRAS.358..113L} {358, 113}
	
	\bibitem[\protect\citeauthoryear{{Lyubarsky}}{{Lyubarsky}}{2014}]{Lyubarsky14}
	{Lyubarsky} Y.,  2014, \mn@doi [\mnras] {10.1093/mnrasl/slu046}, \href
	{https://ui.adsabs.harvard.edu/abs/2014MNRAS.442L...9L} {442, L9}
	
	\bibitem[\protect\citeauthoryear{{Lyutikov} \& {Popov}}{{Lyutikov} \&
		{Popov}}{2020}]{Lyutikov2020}
	{Lyutikov} M.,  {Popov} S.,  2020, arXiv e-prints, \href
	{https://ui.adsabs.harvard.edu/abs/2020arXiv200505093L} {p. arXiv:2005.05093}
	
	\bibitem[\protect\citeauthoryear{{Macquart} \& {Koay}}{{Macquart} \&
		{Koay}}{2013}]{MacquartKoay13}
	{Macquart} J.-P.,  {Koay} J.~Y.,  2013, \mn@doi [\apj]
	{10.1088/0004-637X/776/2/125}, \href
	{https://ui.adsabs.harvard.edu/abs/2013ApJ...776..125M} {776, 125}
	
	\bibitem[\protect\citeauthoryear{{Majid}, {Pearlman}, {Nimmo}, {Hessels},
		{Prince}, {Naudet}, {Kocz}  \& {Horiuchi}}{{Majid} et~al.}{2020}]{Majid2020}
	{Majid} W.~A.,  {Pearlman} A.~B.,  {Nimmo} K.,  {Hessels} J. W.~T.,  {Prince}
	T.~A.,  {Naudet} C.~J.,  {Kocz} J.,   {Horiuchi} S.,  2020, \mn@doi [\apjl]
	{10.3847/2041-8213/ab9a4a}, \href
	{https://ui.adsabs.harvard.edu/abs/2020ApJ...897L...4M} {897, L4}
	
	\bibitem[\protect\citeauthoryear{{Marcote} et~al.,}{{Marcote}
		et~al.}{2017}]{Marcote2017}
	{Marcote} B.,  et~al., 2017, \mn@doi [\apjl] {10.3847/2041-8213/834/2/L8},
	\href {https://ui.adsabs.harvard.edu/abs/2017ApJ...834L...8M} {834, L8}
	
	\bibitem[\protect\citeauthoryear{{Margalit}, {Beniamini}, {Sridhar}  \&
		{Metzger}}{{Margalit} et~al.}{2020a}]{MBSM2020}
	{Margalit} B.,  {Beniamini} P.,  {Sridhar} N.,   {Metzger} B.~D.,  2020a, arXiv
	e-prints, \href {https://ui.adsabs.harvard.edu/abs/2020arXiv200505283M} {p.
		arXiv:2005.05283}
	
	\bibitem[\protect\citeauthoryear{{Margalit}, {Metzger}  \& {Sironi}}{{Margalit}
		et~al.}{2020b}]{Margalit+20}
	{Margalit} B.,  {Metzger} B.~D.,   {Sironi} L.,  2020b, \mn@doi [\mnras]
	{10.1093/mnras/staa1036}, \href
	{https://ui.adsabs.harvard.edu/abs/2020MNRAS.494.4627M} {494, 4627}
	
	\bibitem[\protect\citeauthoryear{{Mereghetti} et~al.,}{{Mereghetti}
		et~al.}{2020}]{Mereghetti+20}
	{Mereghetti} S.,  et~al., 2020, \mn@doi [\apjl] {10.3847/2041-8213/aba2cf},
	\href {https://ui.adsabs.harvard.edu/abs/2020ApJ...898L..29M} {898, L29}
	
	\bibitem[\protect\citeauthoryear{{Metzger}, {Berger}  \& {Margalit}}{{Metzger}
		et~al.}{2017}]{Metzger+17}
	{Metzger} B.~D.,  {Berger} E.,   {Margalit} B.,  2017, \mn@doi [\apj]
	{10.3847/1538-4357/aa633d}, \href
	{https://ui.adsabs.harvard.edu/abs/2017ApJ...841...14M} {841, 14}
	
	\bibitem[\protect\citeauthoryear{{Metzger}, {Margalit}  \& {Sironi}}{{Metzger}
		et~al.}{2019}]{Metzger+19}
	{Metzger} B.~D.,  {Margalit} B.,   {Sironi} L.,  2019, \mn@doi [\mnras]
	{10.1093/mnras/stz700}, \href
	{https://ui.adsabs.harvard.edu/abs/2019MNRAS.485.4091M} {485, 4091}
	
	\bibitem[\protect\citeauthoryear{{Michilli} et~al.,}{{Michilli}
		et~al.}{2018}]{Michilli+18}
	{Michilli} D.,  et~al., 2018, \mn@doi [\nat] {10.1038/nature25149}, \href
	{https://ui.adsabs.harvard.edu/abs/2018Natur.553..182M} {553, 182}
	
	\bibitem[\protect\citeauthoryear{{Murase}, {Kashiyama}  \&
		{M{\'e}sz{\'a}ros}}{{Murase} et~al.}{2016}]{Murase2016}
	{Murase} K.,  {Kashiyama} K.,   {M{\'e}sz{\'a}ros} P.,  2016, \mn@doi [\mnras]
	{10.1093/mnras/stw1328}, \href
	{https://ui.adsabs.harvard.edu/abs/2016MNRAS.461.1498M} {461, 1498}
	
	\bibitem[\protect\citeauthoryear{{Neronov} \& {Vovk}}{{Neronov} \&
		{Vovk}}{2010}]{NV2010}
	{Neronov} A.,  {Vovk} I.,  2010, \mn@doi [Science] {10.1126/science.1184192},
	\href {https://ui.adsabs.harvard.edu/abs/2010Sci...328...73N} {328, 73}
	
	\bibitem[\protect\citeauthoryear{{Os{\l}owski} et~al.,}{{Os{\l}owski}
		et~al.}{2019}]{Oslowski2019}
	{Os{\l}owski} S.,  et~al., 2019, \mn@doi [\mnras] {10.1093/mnras/stz1751},
	\href {https://ui.adsabs.harvard.edu/abs/2019MNRAS.488..868O} {488, 868}
	
	\bibitem[\protect\citeauthoryear{{Pearlman} et~al.,}{{Pearlman}
		et~al.}{2020}]{Pearlman2020}
	{Pearlman} A.~B.,  et~al., 2020, arXiv e-prints, \href
	{https://ui.adsabs.harvard.edu/abs/2020arXiv200508410P} {p. arXiv:2005.08410}
	
	\bibitem[\protect\citeauthoryear{{P{\'e}tri} \& {Lyubarsky}}{{P{\'e}tri} \&
		{Lyubarsky}}{2007}]{Petri2007}
	{P{\'e}tri} J.,  {Lyubarsky} Y.,  2007, \mn@doi [\aap]
	{10.1051/0004-6361:20066981}, \href
	{https://ui.adsabs.harvard.edu/abs/2007A&A...473..683P} {473, 683}
	
	\bibitem[\protect\citeauthoryear{{Petroff} et~al.,}{{Petroff}
		et~al.}{2016}]{Petroff2016}
	{Petroff} E.,  et~al., 2016, \mn@doi [\pasa] {10.1017/pasa.2016.35}, \href
	{https://ui.adsabs.harvard.edu/abs/2016PASA...33...45P} {33, e045}
	
	\bibitem[\protect\citeauthoryear{{Plotnikov} \& {Sironi}}{{Plotnikov} \&
		{Sironi}}{2019}]{Plotnikov&Sironi19}
	{Plotnikov} I.,  {Sironi} L.,  2019, \mn@doi [\mnras] {10.1093/mnras/stz640},
	\href {https://ui.adsabs.harvard.edu/abs/2019MNRAS.485.3816P} {485, 3816}
	
	\bibitem[\protect\citeauthoryear{{Ravi}}{{Ravi}}{2019a}]{Ravi2019b}
	{Ravi} V.,  2019a, \mn@doi [Nature Astronomy] {10.1038/s41550-019-0831-y},
	\href {https://ui.adsabs.harvard.edu/abs/2019NatAs...3..928R} {3, 928}
	
	\bibitem[\protect\citeauthoryear{{Ravi}}{{Ravi}}{2019b}]{Ravi2019}
	{Ravi} V.,  2019b, \mn@doi [\mnras] {10.1093/mnras/sty1551}, \href
	{https://ui.adsabs.harvard.edu/abs/2019MNRAS.482.1966R} {482, 1966}
	
	\bibitem[\protect\citeauthoryear{{Ravi} et~al.,}{{Ravi} et~al.}{2019}]{Ravi+19}
	{Ravi} V.,  et~al., 2019, \mn@doi [\nat] {10.1038/s41586-019-1389-7}, \href
	{https://ui.adsabs.harvard.edu/abs/2019Natur.572..352R} {572, 352}
	
	\bibitem[\protect\citeauthoryear{{Ridnaia} et~al.,}{{Ridnaia}
		et~al.}{2020}]{Ridania2020}
	{Ridnaia} A.,  et~al., 2020, arXiv e-prints, \href
	{https://ui.adsabs.harvard.edu/abs/2020arXiv200511178R} {p. arXiv:2005.11178}
	
	\bibitem[\protect\citeauthoryear{{Sari} \& {Piran}}{{Sari} \&
		{Piran}}{1997}]{SariPiran1997}
	{Sari} R.,  {Piran} T.,  1997, \mn@doi [\apj] {10.1086/304428}, \href
	{https://ui.adsabs.harvard.edu/abs/1997ApJ...485..270S} {485, 270}
	
	\bibitem[\protect\citeauthoryear{{Shannon} et~al.,}{{Shannon}
		et~al.}{2018}]{Shannon2018}
	{Shannon} R.~M.,  et~al., 2018, \mn@doi [\nat] {10.1038/s41586-018-0588-y},
	\href {https://ui.adsabs.harvard.edu/abs/2018Natur.562..386S} {562, 386}
	
	\bibitem[\protect\citeauthoryear{{Spitler} et~al.,}{{Spitler}
		et~al.}{2014}]{Spitler2014}
	{Spitler} L.~G.,  et~al., 2014, \mn@doi [\apj] {10.1088/0004-637X/790/2/101},
	\href {https://ui.adsabs.harvard.edu/abs/2014ApJ...790..101S} {790, 101}
	
	\bibitem[\protect\citeauthoryear{{Tavani} et~al.,}{{Tavani}
		et~al.}{2020}]{Tavani2020}
	{Tavani} M.,  et~al., 2020, arXiv e-prints, \href
	{https://ui.adsabs.harvard.edu/abs/2020arXiv200512164T} {p. arXiv:2005.12164}
	
	\bibitem[\protect\citeauthoryear{{Tavecchio}, {Ghisellini}, {Foschini},
		{Bonnoli}, {Ghirlanda}  \& {Coppi}}{{Tavecchio} et~al.}{2010}]{Tavecchio2010}
	{Tavecchio} F.,  {Ghisellini} G.,  {Foschini} L.,  {Bonnoli} G.,  {Ghirlanda}
	G.,   {Coppi} P.,  2010, \mn@doi [\mnras] {10.1111/j.1745-3933.2010.00884.x},
	\href {https://ui.adsabs.harvard.edu/abs/2010MNRAS.406L..70T} {406, L70}
	
	\bibitem[\protect\citeauthoryear{{Tendulkar} et~al.,}{{Tendulkar}
		et~al.}{2017}]{Tendulkar+17}
	{Tendulkar} S.~P.,  et~al., 2017, \mn@doi [\apjl] {10.3847/2041-8213/834/2/L7},
	\href {https://ui.adsabs.harvard.edu/abs/2017ApJ...834L...7T} {834, L7}
	
	\bibitem[\protect\citeauthoryear{{The CHIME/FRB Collaboration} et~al.,}{{The
			CHIME/FRB Collaboration} et~al.}{2020}]{CHIME2020}
	{The CHIME/FRB Collaboration} et~al., 2020, arXiv e-prints, \href
	{https://ui.adsabs.harvard.edu/abs/2020arXiv200510324T} {p. arXiv:2005.10324}
	
	\bibitem[\protect\citeauthoryear{{Thompson}}{{Thompson}}{2019}]{Thompson2019}
	{Thompson} C.,  2019, \mn@doi [\apj] {10.3847/1538-4357/aafda3}, \href
	{https://ui.adsabs.harvard.edu/abs/2019ApJ...874...48T} {874, 48}
	
	\bibitem[\protect\citeauthoryear{{Thornton} et~al.,}{{Thornton}
		et~al.}{2013}]{Thornton+13}
	{Thornton} D.,  et~al., 2013, \mn@doi [Science] {10.1126/science.1236789},
	\href {https://ui.adsabs.harvard.edu/abs/2013Sci...341...53T} {341, 53}
	
	\bibitem[\protect\citeauthoryear{{Wadiasingh} \& {Timokhin}}{{Wadiasingh} \&
		{Timokhin}}{2019}]{Wadiasingh2019}
	{Wadiasingh} Z.,  {Timokhin} A.,  2019, \mn@doi [\apj]
	{10.3847/1538-4357/ab2240}, \href
	{https://ui.adsabs.harvard.edu/abs/2019ApJ...879....4W} {879, 4}
	
	\bibitem[\protect\citeauthoryear{{Wang}, {Zhang}, {Chen}  \& {Xu}}{{Wang}
		et~al.}{2019}]{Wang2019}
	{Wang} W.,  {Zhang} B.,  {Chen} X.,   {Xu} R.,  2019, \mn@doi [\apjl]
	{10.3847/2041-8213/ab1aab}, \href
	{https://ui.adsabs.harvard.edu/abs/2019ApJ...876L..15W} {876, L15}
	
	\bibitem[\protect\citeauthoryear{{Xu} \& {Zhang}}{{Xu} \&
		{Zhang}}{2016}]{Xu2016}
	{Xu} S.,  {Zhang} B.,  2016, \mn@doi [\apj] {10.3847/0004-637X/832/2/199},
	\href {https://ui.adsabs.harvard.edu/abs/2016ApJ...832..199X} {832, 199}
	
	\bibitem[\protect\citeauthoryear{{Zhang}}{{Zhang}}{2017}]{Zhang17}
	{Zhang} B.,  2017, \mn@doi [\apjl] {10.3847/2041-8213/aa5ded}, \href
	{https://ui.adsabs.harvard.edu/abs/2017ApJ...836L..32Z} {836, L32}
	
	\makeatother
\end{thebibliography}
\end{document}